\documentstyle[11pt]{article}

\makeatletter
\newcount\@gphlinewidth
\newcount\@eepictcnt
\newdimen\@tempdimc
\@gphlinewidth\@wholewidth \divide\@gphlinewidth 4736

\def\thinlines{\let\@linefnt\tenln \let\@circlefnt\tencirc
    \@wholewidth\fontdimen8\tenln \@halfwidth .5\@wholewidth
    \@gphlinewidth\@wholewidth \divide\@gphlinewidth 4736\relax}
\def\thicklines{\let\@linefnt\tenlnw \let\@circlefnt\tencircw
    \@wholewidth\fontdimen8\tenlnw \@halfwidth .5\@wholewidth
    \@gphlinewidth\@wholewidth \divide\@gphlinewidth 4736
    \advance\@gphlinewidth\@ne   
    \relax}
\newif\if@nodotdef \global\@nodotdeftrue
\def\dottedjoin{\global\@jointhemtrue \global\@joinkind=0\relax
  \bgroup\@ifnextchar[{\global\@nodotdeffalse\@idottedjoin}%
                      {\global\@nodotdeftrue\@idottedjoin[\@empty]}}
\long\def\jput(#1,#2)#3{\@killglue\raise#2\unitlength\hbox to \z@{\hskip
#1\unitlength #3\hss}%
\if@jointhem \if@firstpoint \gdef\x@one{#1} \gdef\y@one{#2} \global\@firstpointfalse
 \else\ifcase\@joinkind
    \if@nodotdef
        \@spdottedline{\dotgap@join\unitlength}%
(\x@one\unitlength ,\y@one\unitlength)(#1\unitlength,#2\unitlength)
    \else
	\@dottedline[\dotchar@join]{\dotgap@join\unitlength}%
(\x@one\unitlength,\y@one\unitlength)(#1\unitlength,#2\unitlength)
    \fi
	\or\@dashline[\dashlinestretch]{\dashlen@join\unitlength}[\dotgap@join]%
(\x@one,\y@one)(#1,#2)
	\else\@drawline[\drawlinestretch](\x@one,\y@one)(#1,#2)\fi
    \gdef\x@one{#1}\gdef\y@one{#2}%
 \fi
\fi\ignorespaces}
\def\dottedline{\@ifnextchar [{\@idottedline}{\@ispdottedline}}
\def\@ispdottedline#1(#2,#3){\@ifnextchar (%
{\@iispdottedline{#1}(#2,#3)}{\relax}}
\def\@iispdottedline#1(#2,#3)(#4,#5){\@spdottedline{#1\unitlength}%
(#2\unitlength,#3\unitlength)(#4\unitlength,#5\unitlength)%
\@ispdottedline{#1}(#4,#5)}
\def\@spdottedline#1(#2,#3)(#4,#5){%
    \@tempcnta \@gphlinewidth\relax
    \advance\@tempcnta by 2     
    \special{pn \the\@tempcnta}%
    \@tempdimc=#2\relax
    \@tempcnta \@tempdimc\relax \advance\@tempcnta 2368 \divide\@tempcnta 4736
    \@tempdimc=#3\relax
    \@tempcntb -\@tempdimc\relax\advance\@tempcntb -2368 \divide\@tempcntb 4736
    \@paspecial{\the\@tempcnta}{\the\@tempcntb}%
    \@tempdimc=#4\relax
    \@tempcnta \@tempdimc\relax \advance\@tempcnta 2368 \divide\@tempcnta 4736
    \@tempdimc=#5\relax
    \@tempcntb -\@tempdimc\relax\advance\@tempcntb -2368 \divide\@tempcntb 4736
    \@paspecial{\the\@tempcnta}{\the\@tempcntb}%
    \@tempdimc=#1\relax
    \@tempcnta \@tempdimc\relax \advance\@tempcnta 2368 \divide\@tempcnta 4736
    \@tempcntb \@tempcnta\relax \divide\@tempcntb 1000
    \multiply \@tempcntb 1000 \advance\@tempcnta -\@tempcntb
    \divide\@tempcntb 1000
    \ifnum\@tempcnta < 10
        \special{dt \the\@tempcntb.00\the\@tempcnta}%
    \else\ifnum\@tempcnta < 100
        \special{dt \the\@tempcntb.0\the\@tempcnta}%
    \else
        \special{dt \the\@tempcntb.\the\@tempcnta}%
    \fi\fi
    \ignorespaces
}
\def\@iiidashline[#1]#2[#3](#4,#5)(#6,#7){%
\@dashline[#1]{#2\unitlength}[#3](#4,#5)(#6,#7)%
\@iidashline[#1]{#2}[#3](#6,#7)}
\long\def\@dashline[#1]#2[#3](#4,#5)(#6,#7){{%
\x@diff=#6\unitlength \advance\x@diff by -#4\unitlength
\y@diff=#7\unitlength \advance\y@diff by -#5\unitlength
\@tempdima=#2\relax \advance\@tempdima -\@wholewidth
\sqrtandstuff{\x@diff}{\y@diff}{\@tempdima}%
\ifnum\num@segments <3 \num@segments=3\fi
\@tempdima=\x@diff \@tempdimb=\y@diff
\divide\@tempdimb by\num@segments
\divide\@tempdima by\num@segments
{\ifx#3\@empty \relax
    \ifdim\@tempdima < 0pt \x@diff=-\@tempdima\else\x@diff=\@tempdima\fi
    \ifdim\@tempdimb < 0pt \y@diff=-\@tempdimb\else\y@diff=\@tempdimb\fi
    \global\setbox\@dotbox\hbox{%
                \@absspdrawline(0pt,0pt)(\@tempdima,\@tempdimb)}%
    \else\global\setbox\@dotbox\hbox{%
        \@spdottedline{#3\unitlength}(0pt,0pt)(\@tempdima,\@tempdimb)}%
    \fi}%
\advance\x@diff by -\@tempdima 
\advance\y@diff by -\@tempdimb
\@tempdima=\num@segments\@wholewidth \@tempdima=2\@tempdima
\@tempcnta\@tempdima\relax \@tempdima=#2\relax \@tempdimb=0.5\@tempdima
\@tempcntb\@tempdimb\relax \advance\@tempcnta by \@tempcntb 
\divide\@tempcnta by\@tempdima \advance\num@segments by -\@tempcnta
\ifnum #1=0 \relax\else\ifnum #1 < -100
  \typeout{***dashline: reduction > -100 percent implies blankness!***}
\else\num@segmentsi=#1 \advance\num@segmentsi by 100
     \multiply\num@segments by\num@segmentsi \divide\num@segments by 100
\fi\fi
\divide\num@segments by 2 
\ifnum\num@segments >0 
 \divide\x@diff by\num@segments
 \divide\y@diff by\num@segments
 \advance\num@segments by\@ne 
 \else\num@segments=2 
\fi
\@xdim=#4\unitlength \@ydim=#5\unitlength
\@killglue
\loop \ifnum\num@segments > 0
\unskip\raise\@ydim\hbox to\z@{\hskip\@xdim \copy\@dotbox\hss}%
\advance\num@segments \m@ne\advance\@xdim\x@diff\advance\@ydim\y@diff%
\repeat}%
\ignorespaces}
\def\@drawline[#1](#2,#3)(#4,#5){{%
\@drawitfalse\@horvlinefalse
\ifnum#1 <0 \relax\else\@horvlinetrue\fi
\if@horvline
 \@spdrawline(#2,#3)(#4,#5)
\else\@drawittrue\fi
\if@drawit
\ifnum #1=0 \relax \else\ifnum #1 < -99
  \typeout{***drawline: reduction <= -100 percent implies blankness!***}%
\else\num@segmentsi=#1 \advance\num@segmentsi by 50
     \multiply\num@segmentsi 2
\fi\fi
\@dashline[\num@segmentsi]{10pt}[\@empty](#2,#3)(#4,#5)
\fi}\ignorespaces}
\def\@spdrawline(#1,#2)(#3,#4){%
   \@absspdrawline(#1\unitlength,#2\unitlength)(#3\unitlength,#4\unitlength)
   \ignorespaces
}
\def\@absspdrawline(#1,#2)(#3,#4){%
    \special{pn \the\@gphlinewidth}%
    \@tempdimc=#1\relax
    \@tempcnta \@tempdimc\relax \advance\@tempcnta 2368 \divide\@tempcnta 4736
    \@tempdimc=#2\relax
    \@tempcntb -\@tempdimc\relax \advance\@tempcntb -2368 \divide\@tempcntb 4736
    \@paspecial{\the\@tempcnta}{\the\@tempcntb}%
    \@tempdimc=#3\relax
    \@tempcnta\@tempdimc\relax \advance\@tempcnta 2368 \divide\@tempcnta 4736
    \@tempdimc=#4\relax
    \@tempcntb -\@tempdimc\relax \advance\@tempcntb -2368 \divide\@tempcntb 4736
    \@paspecial{\the\@tempcnta}{\the\@tempcntb}%
    \special{fp}%
    \ignorespaces
}
\def\@paspecial#1#2{%
    \special{pa #1 #2}%
}
\def\Thicklines{\let\@linefnt\tenlnw \let\@circlefnt\tencircw
    \@wholewidth\fontdimen8\tenlnw \@wholewidth 1.5\@wholewidth
    \@halfwidth .5\@wholewidth
    \@gphlinewidth\@wholewidth \divide\@gphlinewidth 4736\relax}
\def\@circlespecial#1#2#3#4{%
	      \special{pn \the\@gphlinewidth}%
	      \special{ar 0 0 #1 #2 #3 #4}
}
\def\@arc#1#2#3#4{%
	\@tempdima #1\unitlength
	\@tempdimb #2\unitlength
        \@tempcnta\@tempdima \advance\@tempcnta 4736 \divide\@tempcnta 9473
	\@tempcntb\@tempdimb \advance\@tempcntb 4736 \divide\@tempcntb 9473
	\setbox\@tempboxa\hbox{%
	    \@circlespecial{\the\@tempcnta}{\the\@tempcntb}{#3}{#4}}%
        \wd\@tempboxa\z@ \box\@tempboxa}
\def\circle{%
    \@ifstar{\copy\@filltype\@circle}{\@circle}}
\def\@circle#1{\@arc{#1}{#1}{0}{6.2832}}
\def\ellipse{%
    \@ifstar{\copy\@filltype\@ellipse}{\@ellipse}}
\def\@ellipse#1#2{{\@arc{#1}{#2}{0}{6.2832}}}
\def\arc#1#2#3{\@arc{#1}{#1}{#2}{#3}}
\def\@linespecial#1#2{%
	      \special{pn \the\@gphlinewidth}%
	      \special{pa 0 0}%
	      \special{pa #1 #2}%
	      \special{fp}%
}
\def\line(#1,#2)#3{\@xarg #1\relax \@yarg #2\relax
\@linelen=#3\unitlength
\ifnum\@xarg =0 \@vline 
  \else \ifnum\@yarg =0 \@hline \else \@ssline\fi
\fi}
\def\@ssline{%
	\ifnum\@xarg< 0
	  \@negargtrue \@xarg -\@xarg \@tempdima -\@linelen
	\else
	  \@negargfalse \@tempdima\@linelen
	\fi
	\@tempcnta\@linelen \divide\@tempcnta 4736
        \@yyarg -\@yarg \multiply\@yyarg \@tempcnta \divide\@yyarg\@xarg
 	\if@negarg
	    \@tempcnta -\@tempcnta
	\fi
	\setbox\@linechar\hbox{\@linespecial{\the\@tempcnta}{\the\@yyarg}}%
	\wd\@linechar\@tempdima
	\@clnht\@linelen
        \multiply\@clnht\@yarg
        \divide\@clnht\@xarg
	\ifnum\@yarg< 0
	  \@clnht -\@clnht
	  \ht\@linechar\z@ \dp\@linechar\@clnht
	\else
	  \ht\@linechar\@clnht \dp\@linechar\z@
	\fi
	\box\@linechar
}
\def\@sline{%
	\@ssline
	\if@negarg
	  \@yyarg -\@yarg
	\else
	  \@yyarg \@yarg
	\fi
	\setbox\@linechar\hbox{\@linefnt\@getlinechar(\@xarg,\@yyarg)}%
	\ifnum\@yarg> 0
	  \let\@upordown\raise
	  \advance\@clnht -\ht\@linechar
	\else
	  \let\@upordown\lower
	\fi
	\if@negarg \kern\wd\@linechar \fi
}
\def\spline(#1,#2){%
    \special{pn \the\@gphlinewidth}%
    \@spline(#1,#2)%
}
\def\@spline(#1,#2){%
    \@tempdima #1\unitlength
    \@tempdimb #2\unitlength
    \@tempcnta \@tempdima \advance\@tempcnta 2368 \divide\@tempcnta 4736
    \@tempcntb -\@tempdimb \advance\@tempcntb -2368 \divide\@tempcntb 4736
    \@paspecial{\the\@tempcnta}{\the\@tempcntb}%
    \@ifnextchar ({\@spline}{\special{sp}}%
}
\def\path(#1,#2){%
    \special{pn \the\@gphlinewidth}%
    \@path(#1,#2)%
}
\def\@path(#1,#2){%
    \@tempdima #1\unitlength
    \@tempdimb #2\unitlength
    \@tempcnta \@tempdima \advance\@tempcnta 2368 \divide\@tempcnta 4736
    \@tempcntb -\@tempdimb \advance\@tempcntb -2368 \divide\@tempcntb 4736
    \@paspecial{\the\@tempcnta}{\the\@tempcntb}%
    \@ifnextchar ({\@path}{\special{fp}}%
}

\newdimen\maxovaldiam \maxovaldiam 40pt\relax

\def\@oval(#1,#2)[#3]{\begingroup\boxmaxdepth \maxdimen
  \@ovttrue \@ovbtrue \@ovltrue \@ovrtrue
  \@tfor\@tempa :=#3\do{\csname @ov\@tempa false\endcsname}\@ovxx
  #1\unitlength \@ovyy #2\unitlength
  \@tempdimb \ifdim \@ovyy >\@ovxx \@ovxx\else \@ovyy \fi
  \@ovro \ifdim \@tempdimb>\maxovaldiam \maxovaldiam\else\@tempdimb\fi\relax
  \divide \@ovro \tw@
  \@ovdx\@ovxx \divide\@ovdx \tw@
  \@ovdy\@ovyy \divide\@ovdy \tw@
  \setbox\@tempboxa \hbox{%
  \if@ovr \@ovverta\fi
  \if@ovl \kern \@ovxx \@ovvertb\kern -\@ovxx \fi
  \if@ovt \@ovhorz \kern -\@ovxx \fi
  \if@ovb \raise \@ovyy \@ovhorz \fi}
  \ht\@tempboxa\z@ \dp\@tempboxa\z@
  \@put{-\@ovdx}{-\@ovdy}{\box\@tempboxa}%
  \endgroup}

\def\@qcirc#1#2#3#4{%
    \special{pn \the\@gphlinewidth}%
    \@eepictcnt \@gphlinewidth \divide\@eepictcnt 2
    \@tempcnta #1 
      \advance\@tempcnta 2368 \divide\@tempcnta 4736
      \advance\@tempcnta\@eepictcnt
    \@tempcntb #2 \divide\@tempcntb 4736 \advance\@tempcntb 2
    \hbox{%
        \@qcircspecial{\the\@tempcnta}{-\the\@eepictcnt}{\the\@tempcntb}{#3}{#4}}%
}
\def\@qcircspecial#1#2#3#4#5{\special{ar #1 #2 #3 #3 #4 #5}}

\def\@ovverta{\vbox to \@ovyy{%
    \if@ovb
        \kern \@ovro
        \@qcirc{\@ovro}{\@ovro}{3.1416}{4.7124}\nointerlineskip
    \else
        \kern \@ovdy
    \fi
    \leaders\vrule width \@wholewidth\vfil \nointerlineskip
    \if@ovt
        \@qcirc{\@ovro}{\@ovro}{1.5708}{3.1416}\nointerlineskip
        \kern \@ovro
    \else
        \kern \@ovdy
    \fi
}\kern -\@wholewidth}

\def\@ovvertb{\vbox to \@ovyy{%
    \if@ovb
        \kern \@ovro
        \@qcirc{-\@ovro}{\@ovro}{4.6124}{6.2832}\nointerlineskip
    \else
        \kern \@ovdy
    \fi
    \leaders\vrule width \@wholewidth\vfil \nointerlineskip
    \if@ovt
        \@qcirc{-\@ovro}{\@ovro}{0}{1.6708}\nointerlineskip
        \kern \@ovro
    \else
        \kern \@ovdy
    \fi
}\kern -\@wholewidth}

\def\@ovhorz{\hbox to \@ovxx{%
    \if@ovr \kern \@ovro\else \kern \@ovdx \fi
    \leaders \hrule height \@wholewidth \hfil
    \if@ovl \kern \@ovro\else \kern \@ovdx \fi
    }}

\def\allinethickness#1{\let\@linefnt\tenlnw \let\@circlefnt\tencircw
    \@wholewidth #1 \@halfwidth .5\@wholewidth
    \@gphlinewidth\@wholewidth \divide\@gphlinewidth 4736\relax}

\newbox\@filltype
\setbox\@filltype\hbox{\special{bk}}
\def\filltype#1{\@nameuse{ft@#1}}
\def\ft@black{\setbox\@filltype\hbox{\special{bk}}}
\def\ft@white{\setbox\@filltype\hbox{\special{wh}}}
\def\ft@shade{\setbox\@filltype\hbox{\special{sh}}}
\makeatother

\def\beq{\begin{equation}}
\def\eeq{\end{equation}}
\def\bea{\begin{eqnarray}}
\def\eea{\end{eqnarray}}

\def\nn{\nonumber}
\def\ox{\otimes}
\def\ld{\lambda}
\def\Yq{Y_{Lq}}
\def\Yl{Y_{Ll}}

\title
{\LARGE\bf Effective Lagrangian \\ for a Technicolor Model\\         
without\\  Exact Custodial Symmetry }
\bigskip 
\author{T.Yoshikawa, H.Takata  and  T.Morozumi \\
                             \\ 
\normalsize \em Department of Physics, Hiroshima University\\ 
\normalsize \em 1-3-1 Kagamiyama, Higashi-Hiroshima, 724\\
\normalsize \em Japan} 
 
\date{}
\begin{document}
\setlength{\baselineskip}{24pt}
\maketitle
\begin{picture}(0,0)
\put(325,240){HUPD-9406}
\put(325,225){hep-ph/9403343}
\put(325,210){March, 1994}
\end{picture}
\vspace{-24pt}

\begin{abstract} \normalsize\noindent
Effective Lagrangian including technimesons is constructed
 for a realistic one-family Technicolor model without 
exact custodial symmetry. Tree level contribution to
oblique correction parameters $S$ and $U$ due to spin 1
technimesons are computed with the effective
Lagrangian.
 An isospin breaking term
 which is associated with technilepton vector mesons
 gives a negative contribution to the electroweak radiative
correction parameter $S$ due to mixing between $I=0$ and $I=1$ vector 
mesons. $U$ receives non-zero contribution due to exotic left-handed
charged vector mesons and its sign can be both negative and positive.
\end{abstract}

\section{Introduction}
It has been shown that Technicolor models are strongly constrained by
precision measurements of electorweak parameters.
In particular, QCD scale up one-family
Technicolor model with exact custodial symmetry
seem to be already
excluded by studying an oblique correction parameter $S$ \cite{pes}. 
\bea
       & & S_{theory} = 0.28 \times 4 = 1.1  \nn \\
       & & S_{exp} = - 0.42 \pm 0.45 \nn
\eea
where $S_{theory}$ is an estimation with vector and axial
vector meson dominance assumption for QCD scale up
one-family Technicolor model. (See appendix E.)
A factor of 4 comes from the fact that the model contains 4
$SU(2)$ doublet of technifermions.
$S_{exp}$ is quoted from \cite{sexp} and the reference point
of the standard model 
is taken at $m_t = 150 GeV$ and $m_H = 1 TeV$. 
However, according to ref.\cite{apel}, this is not the
case for Technicolor models without exact custodial
symmetry. A realistic model is proposed for a one-family 
Technicolor model. In their model, isospin breaking is
introduced for a light technilepton doublet. The doublet
contributes to a radiative correction parameter $S$ in
negative sign while keeping $\rho $ parameter nearly equal to 1.
In their analysis, free 
technifermion model is used to compute $S$ parameter.

In this paper, we compute the oblique corrections ($S$, $T$
and $U$ ) in an 
alternative way. We construct an effective Lagrangian with
low lying technimesons for the one-family Technicolor model
without exact custodial symmetry. By using the effective
Lagrangian, we can compute non-perturbative effect  
on the radiative correction
parameters S,T, and U due to
bound states of technifermions. 
Our paper is organized in the following
way. In \S 2, we review feature of the model.
With some assumption on the low lying technimesons' 
spectrum,
 we construct a low energy effective Lagrangian. In \S 3, 
$S$, $T$, and $U$ parameters are computed with the Lagrangian.
It is shown that S parameter receives negative contribution 
due to the mixing between isosinglet and isotriplet
techni-vector mesons.  \S 4 is devoted to finding the 
range of the paramerters for negative S.
Conclusions and discussion are 
summarized in \S 5.

\section{ Effective Lagrangian for a Technicolor
 model without exact custodial symmetry }
Let us describe the model briefly \cite{apel}. The model has a
global symmetry : $ G = SU(6)_L \ox SU(6)_R \ox SU(2)_L \ox
U(1)_{2R} \ox U(1)_{8L} \ox U(1)_{8R} \ox U(1)_V $ which 
is spontaneously broken to $ H = SU(6)_V \ox U(1)_{2V} \ox
U(1)_{8V} \ox U(1)_V $. The part of G must be broken
explicitly in order that unnecessary massless physical 
Nambu-Goldstone bosons disappear.

The technifermions are assigned to the following representations
of $SU(3)_c \ox SU(2) \ox U(1)_Y \ox SU(N_{TC})$:
\bea
  ( U, D )_L &=& ( 3, 2, \frac{\Yq}{2}, N_{TC} ), \nn \\
      U_R   &=& ( 3, 1, \frac{\Yq}{2}+\frac{1}{2}, N_{TC} ), \nn \\ 
      D_R   &=& ( 3, 1, \frac{\Yq}{2}-\frac{1}{2}, N_{TC} ), \nn \\
  ( N, E )_L &=& ( 1, 2,\frac{\Yl}{2}+\frac{1}{2}, N_{TC} ),\nn \\
      E_R   &=& ( 1, 1, \frac{\Yl}{2}-\frac{1}{2}, N_{TC} ), \nn 
\eea
where $\Yq(\Yl)$ is hypercharge of lefthanded
techniquark(technilepton) ($\Yq = 1/3$, $\Yl = - 1$ ). 

The following mass spectrum is assumed for technifermions:
\medskip

(1) $M_U = M_D $,  	      

(2) $M_N < M_E < M_U $. 

\medskip
\def\CL{{\calL}}
$SU(6)_V$ symmetry is preserved because techniquarks are degenerate,
while $SU(2)_V$ symmetry is explicitly broken due to isospin breaking 
of technileptons. 
To proceed further, we need to know the technimesons spectrum 
of the model. Since the model does not have the same global 
symmetry as that of QCD ( $SU(2)_L \ox SU(2)_R$ in chiral
limit ) and underlying dynamics may be also different 
from that, we can not simply scale up the mesons' spectrum in QCD. 
Here we make use of the global symmetry as a guide to construct
an effective Lagrangian. Global symmetry strongly constraints
the structure of effective Lagrangian as well as properties of bound
states included in effective Lagrangian.
Concerned with technimesons, which are bound states of technifermions,
we need to make a few assumptions.  In this paper, we include 
only spin 0 (Nambu Goldstone Bosons (NGB) and Pseudo Nambu Goldstone
Bosons (PNGB)) and spin 1 mesons. For the purpose of studying 
tree level contribution to the oblique correction
parameters, S,T, 
and U, other mesons with higher spins (spin$\ge 2$)
can be ignored because they do not contribute to self energy corrections
of gauge bosons. Further we only keep $O(p^2)$ terms for NGB and PNGB
sector and ignore their loop effects and $O(p^4)$ counter terms.
About spin 1 mesons, we employ the approach of including vector mesons 
into chiral Lagrangian \cite{wein}\cite{ban} and extend it to our case.
In table(1), technimesons 
and their technifermion contents
as well as their J,P,C, and I are listed.(For spin 0 sector, we quote them
from ref.\cite{apel}.) Note that charged technilepton NGBs
($\Pi^\pm$) do not have definite parity because $\Pi^\pm$
are NGBs associated with $SU(2)_L$ not $SU(2)_A$. ( Note
that we do not have full $SU(2)_R$ symmetry.) In the same
way, exotic left-handed charged vector mesons($A_L$) are
introduced so that they interact with
$\Pi^\pm$ and  $SU(2) \ox U(1)$ gauge bosons etc without
loss of the invariance. 
Corresponding to spin$0$, vector, and axial and  left-handed vector mesons, 
the effective Lagrangian consists of three parts:
${\cal L}_{S}$, ${\cal L}_{V}$, and ${\cal L}_{A}$:
The explicit form for them will be presented below.

${\cal L}_S$ ( spin$0$ sector )\\
This part of the Lagrangian  consists of $O(p^2)$ terms of
NGBs and PNGBs.
\bea
 {\cal L}_S &=& F_6^2 tr( \alpha_{6\perp })^2 
          + F_L^2 tr( \alpha_{L\perp })^2 
          + F_2^2 tr( \alpha_{2\perp })^2 
          + F_8^2 tr( \alpha_{8\perp })^2 \nn \\
          &+& \beta_{\perp} tr( \alpha_{8\perp } \tau^3
                                       \alpha_{2\perp} )
          + {\cal L}_{br}^\prime ,
\label{eq:S}
\eea
where,
\bea
 {\alpha_{6\perp }}_\mu
        &=& 2 \sum_{\alpha = 1}^{35}\pmatrix{
                               T^{\alpha} & \ \cr
                                   \ & 0 
                                           }
            tr [ T^{\alpha} \frac{ \xi_6 \nabla_{R\mu} 
                 \xi_6^{\dagger} - \xi_6^{\dagger} 
                 \nabla_{L\mu} \xi_6 }{2i} 
              ],  \\
 {\alpha _{L\perp }}_\mu 
        &=& - 2 \sum_{a = 1}^{2} \frac{1}{2}\pmatrix{
                                    0 & \ \cr
                                    \ & \tau^a
                                            } 
           tr [ \frac{\tau^{a}}{2} \frac{ \xi_2^{\dagger} 
                 {\hat{\nabla}}_{L\mu} \xi_2 }{2i} 
              ],  \\
 {\alpha _{2\perp }}_\mu 
        &=& 2 \frac{1}{2} \pmatrix{
                                0 & \ \cr
                                \ & \tau^3
                                   } 
          tr [ \frac{\tau^{3}}{2} \frac{ \xi_1^{\dagger}
               \hat{\nabla}_{R\mu} \xi_1 - \xi_2^{\dagger}
               \hat{\nabla}_{L\mu} \xi_2 }{2i} 
            ],  \\  
 {\alpha _{8\perp }} _\mu 
        &=& 2 \frac{1}{4 \sqrt{3}} \pmatrix{
                                     I_6 & \ \cr
                                     \ & -3 I_2
                                           }
           [tr\{ \frac{1}{4 \sqrt{3}}  \frac{ \xi_6 \nabla_{R\mu}
             \xi_6^{\dagger} - \xi_6^{\dagger} \nabla_{L\mu}
             \xi_6 }{2i} \} \nn \\  
   & & \hspace{1.4cm}     + 
           tr \{ \frac{-3}{4 \sqrt{3}} 
                \frac{ \xi_1^{\dagger} \hat{\nabla}_{R\mu} \xi_1 - 
                \xi_2^{\dagger} \hat{\nabla}_{L\mu}
                                         \xi_2}{2i} \} 
           ],  
\eea
where, 
\bea 
  \nabla_{R\mu} 
        &=& \partial_\mu 
           + i g^{\prime} (\frac{\tau^3}{2} \ox I_3) B_\mu 
           + i g^{\prime} (\frac{Y_{LQ}}{2} \ox I_3) B_\mu,  
                 \label{eq:6}\\
  \nabla_{L\mu} 
        &=& \partial_\mu 
           + i g \sum_{A=1}^{3}(\frac{\tau^A}{2} \ox I_3) W_\mu^A 
           +  i g^{\prime} (\frac{Y_{LQ}}{2} \ox I_3) B_\mu  ,\\
  \hat{\nabla}_{R\mu} 
        &=& \partial_\mu  
           + i g^{\prime} \frac{\tau^3}{2}B_\mu 
           + i g^{\prime} \frac{Y_{Ll}}{2} B_\mu , \\ 
  \hat{\nabla}_{L\mu} 
        &=& \partial_\mu 
           + i g \sum_{A=1}^{3}\frac{\tau^A}{2} W_\mu^A 
           +  i g^{\prime} \frac{Y_{Ll}}{2}B_\mu  ,
\label{eq:9} 
\eea
\bea
  \xi_6 &=& \exp( i\sum_{\alpha=1}^{35} \frac{T^{\alpha} P^{\alpha}}{F_6}
                  + i\frac{1}{4\sqrt{3}}\frac{\theta_8}{F_8} ),
          \label{eq:10} \\
  \xi_2 &=& \exp( i\sum_{a=1}^{2} \frac{\tau^a \Pi^a}{2 F_L}
             + i\frac{\tau^3 \Pi^3}{2 F_2} 
                  - i\frac{3}{4\sqrt{3}}\frac{\theta_8}{F_8} ),\\
  \xi_1 &=& \exp( - i\frac{\tau^3 \Pi^3 }{2 F_2 } 
                 + i\frac{3}{4\sqrt{3}}\frac{\theta_8}{F_8} ). 
\label{eq:12}
\eea
In (\ref{eq:S}) ${\cal L}_{br}^\prime$ consists of explicit
breaking terms which make physical NGB massive.
Without ${\cal L}_{br}^\prime$ and $SU(2)_L \ox U(1)_Y$
gauge interaction, we have 3 color singlet physical massless
NGBs which are linear combinations of $\theta_8, \Pi^3,
\Pi^\pm, P^3$ and $P^\pm$. Here $P^\pm$ and $P^3$ are $SU(2)$
triplet and color singlet bound states of techniquarks. By
adding ${\cal L}_{br}^\prime$, we can make them massive.
(See appendix D for the details.) In (3), $\tau^a$ is a
projection into $SU(2)_L$ $(a=1,2)$ part.
In (\ref{eq:6})-(\ref{eq:9}), $W^A$ and $B$ are $SU(2)\ox U(1)_Y $ gauge
bosons.
In (\ref{eq:10})-(\ref{eq:12}), $\xi$s 
correspond to NGB fields.  
$ T^{\alpha}$ are generator of $SU(6)$ and $\tau^a$, $\tau^3$ are
Pauli matrices.
$P^{\alpha}$ are 35 NGBs for broken
$SU(6)_A$ symmetry, which are techniquark bound states. 
$\Pi^a(a = 1, 2)$ and $\Pi^3$ are NGBs for
broken $SU(2)_L(\tau^a : a= 1,2 )$ and $U(1)_{2A}$ symmetry respectively,
which are technilepton bound states. $\theta_8$ is a
NGB for $U(1)_{8A}$ symmetry. $F_6, F_L,
F_2$ and $F_8$ are decay constants for these NGBs.  
Note that $F_L$ is not degenerate with  $F_2$  
because custodial symmetry is broken in the sector 
and $\Pi_L$  form an irreducible representation 
under $U(1)_{2V}$
which $\Pi^3$ does not belong to. The difference between
$F_L$ and $F_2$
gives rise to small
deviation of $\rho$ parameter from 1 (see (\ref{eq:T})).  
 The transformation of $\xi$s are given by;
\bea
  \xi_6^{\prime} 
             &=& g_{L6} {\xi}_6 h_6^{\dagger} 
              = h_6 {\xi}_6 g_{R6}^{\dagger},\\
  {\xi}_2^{\prime} 
             &=& g_{L2} {\xi}_2 h_2^{\dagger},\\
  {\xi}_1^{\prime} 
             &=& g_{R2} {\xi}_1 h_2^{\dagger},\\
  \frac{\theta_8^\prime }{F_8} 
             &=& \frac{\theta_8}{F_8} + \phi_{R8} - \phi_{L8},
\eea
where,
\bea
 g_{L6} &=& \exp{( iT_6^\alpha \phi_{L6}^\alpha 
                 + i \frac{1}{4\sqrt{3}} \phi_{L8} 
                 + i \frac{1}{4} \phi_V) },
             \\
 g_{R6} &=& \exp{( iT_6^\alpha \phi_{R6}^\alpha 
                 + i \frac{1}{4\sqrt{3}} \phi_{R8} 
                 + i \frac{1}{4} \phi_V )},  
\eea
\bea
 g_{L2} &=& \exp{( i\sum_{a=1}^2 \tau^a \phi_{L2}^a 
                   + i \tau^3 \phi_{L2}^3 
                   - i \frac{3}{4\sqrt{3}} \phi_{L8} 
                   + i \frac{1}{4} \phi_V )},  \\
 g_{R2} &=& \exp{( 
                    i \tau^3 \phi_{R2}^3 
                  - i \frac{3}{4\sqrt{3}} \phi_{R8} 
                  + i \frac{1}{4} \phi_{V}) },
\eea
\bea
 h_6 &=& \hat{h}_6 h_{Q1}, \\
 \hat{h}_6 &=& \exp{( i T_6^\alpha \phi_{V6}^\alpha )},   \\
 h_{Q1} &=& \exp{(i \frac{1}{2\sqrt{3}} \phi_{Q1} )},
\eea
\bea
 h_2 &=& \hat{h}_2 h_{L1}, \\
 \hat{h}_2 &=& \exp{( i\frac{\tau^3}{2} \phi_{V2} )}, \\
 h_{L1} &=& \exp{( i \frac{1}{2} \phi_{L1} )},
\eea
and
$\phi_{L8}$ and $\phi_{R8}$ are $x$ independent parameters
of $ U(1)_{8L} \ox U(1)_{8R} $ transformations. 
It is not difficult to see that
$\alpha_{6\perp\mu}$ , $\alpha_{L\perp\mu}$ and
$\alpha_{2\perp\mu}$ transform as:
\bea
 \alpha_{6\perp\mu}^\prime &=& \hat{h}_6 
              \alpha_{6\perp\mu} \hat{h}_6^\dagger , \\
 \alpha_{2L\mu}^\prime &=& \hat{h}_2 
               \alpha_{2L\mu} \hat{h}_2^\dagger ,\\
 \alpha_{2\perp\mu}^\prime &=& \alpha_{2\perp\mu} , \\
 \alpha_{8\perp\mu}^\prime &=& \alpha_{8\perp\mu} .
\eea

${\cal L}$ (Vector meson($1^{--}$) sector)\\
In addition to NGBs and PNGBs, we incorporate vector mesons into 
the effective Lagrangian. Corresponding to unbroken
symmetry:  $SU(6)_V \ox U(1)_{2V} \ox U(1)_{8V} \ox U(1)_V
$, we introduce 38 vector mesons, 35 of them are
techniquark bound states which belong to the adjoint representation
of $SU(6)_V$
. Corresponding to $U(1)_{qV}$ ($U(1)_V$ for techniquark sector)
,$U(1)_{lV}$ ($U(1)_V$ for technilepton sector) and $U(1)_{2V}$,
three neutral vector mesons, $techni\ \omega_q$
($\omega_{6\mu}$), 
$techni\ \omega_l$ ($\omega_{2\mu}$), 
and $ techni\ \rho_l$ ($\rho_{2\mu}$)
are introduced.
For the  techniquark sector of the effective lagrangian,
we can just extend the approach of \cite{wein} \cite{ban} 
into the larger symmetry, i.e,
chiral $SU(6)_L\ox SU(6)_R \ox U(1)_{6V}$.
On the otherhand, for technilepton 
 sector, non-trivial isospin breaking terms 
are introduced. Let us record vector meson part of the effective
Lagrangian first,
\bea
 {\cal L}_V &=& \frac{1}{2} tr F_{V_6}^{\mu\nu} {F_{V_6}}_{\mu\nu} 
            + \frac{1}{2} tr F_{V_2}^{\mu\nu} {F_{V_2}}_{\mu\nu}   
            + \frac{1}{2} tr F_{V_{\omega6}}^{\mu\nu} 
                                   {F_{V_{\omega6}}}_{\mu\nu}
            + \frac{1}{2} tr F_{V_{\omega2}}^{\mu\nu} 
                                   {F_{V_{\omega2}}}_{\mu\nu}
               \nn \\  
           &-& M_{V_6}^2 
                  tr( V_{6\mu} - \frac{i}{G_6}\alpha_{6\parallel\mu} )^2
            - M_{V_{\omega6}}^2 
                  tr( V_{\omega6\mu}  
                     -\frac{i}{G_{\omega6}}
                           \alpha_{\omega6\parallel\mu} )^2  \nn \\
          &-& M_{V_2}^2 
                  tr( V_{2\mu} - \frac{i}{G_2}\alpha_{2\parallel\mu} )^2
            - M_{V_{\omega2}}^2 
                  tr( V_{\omega2\mu}  
                     - \frac{i}{G_{\omega2}}
                           \alpha_{\omega2\parallel\mu} )^2 \nn \\
          &+& \alpha_V 
                  tr F_{V_2}^{\mu\nu} \tau^3
                                 {F_{V_{\omega2}}}_{\mu\nu} \nn \\
          &-& \beta_V 
                  tr( V_{2\mu} - \frac{i}{G_2}\alpha_{2\parallel\mu} )
                \tau^3
                    ( V_{\omega2\mu}  
                     -\frac{i}{G_{\omega2}}\alpha_{\omega2\parallel\mu} ) ,
 \label{eq:lv}
\eea
where
\bea
 {\alpha_{6\parallel }}_\mu
        &=& 2 \sum_{\alpha = 1}^{35}\pmatrix{
                              T^{\alpha} & \ \cr
                                \  & 0 
                                             } 
            tr [ T^{\alpha} 
                  \frac{ \xi_6 \nabla_{R\mu} \xi_6^{\dagger} 
                + \xi_6^{\dagger} \nabla_{L\mu} \xi_6 }{2i} 
              ],  \\
 {\alpha _{\omega6\parallel }}_\mu 
        &=& 2 \frac{1}{2\sqrt{3}}\pmatrix{
                               I_6 & \ \cr
                                \ & 0
                                         }
          tr [ \frac{1}{2\sqrt{3}} 
               \frac{ \xi_6 {\nabla}_{R\mu} \xi_6^\dagger 
                   + \xi_6^{\dagger} {\nabla}_{L\mu} \xi_6 }{2i} 
            ],  \\
 {\alpha _{2\parallel }}_\mu 
        &=& 2 \frac{1}{2}\pmatrix{
                               0 & \ \cr
                             \ & \tau^3
                                  } 
          tr [ \frac{\tau^{3}}{2} 
               \frac{ \xi_1^{\dagger} \hat{\nabla}_{R\mu} \xi_1 
                   + \xi_2^{\dagger} \hat{\nabla}_{L\mu} \xi_2 }{2i} 
            ],  \\
 {\alpha _{\omega2\parallel }}_\mu 
        &=& 2 \frac{1}{2}\pmatrix{
                              0 & \ \cr
                              \ & I_2
                                  }
          tr [ \frac{1}{2} 
               \frac{ \xi_1^{\dagger} \hat{\nabla}_{R\mu} \xi_1 
                   + \xi_2^{\dagger} \hat{\nabla}_{L\mu} \xi_2 }{2i} 
            ].  
\eea


Vector mesons are decomposed into their component fields, 
\bea
  V_{6\mu} &=& i\sum_{\alpha=1}^{35} \pmatrix{
                                    T^{\alpha} & \ \cr
                                    \ & 0 
                                         } \rho_{6\mu}^{\alpha }, \\
  V_{\omega6\mu} &=& i\frac{1}{2\sqrt{3}} \pmatrix{
                                    I_6 & \ \cr
                                    \ & 0
                                           } {\omega_{6\mu}}, \\
  V_{2\mu} &=& i\frac{1}{2}\pmatrix{
                                  0 & \ \cr
                                   \ & \tau^3
                                            } \rho_{2\mu}, \\
  V_{\omega2\mu} &=& i\frac{1}{2}\pmatrix{
                                  0 & \ \cr
                                   \ & I_2
                                             } {\omega_{2\mu}}. 
\eea
The quantities defined above transform under G in the
following way,
\bea
   \left\{
       \begin{array}{@{\,}ll} 
           \alpha_{6\parallel\mu}^\prime &= 
                  \hat{h}_6 \alpha_{6\parallel\mu} \hat{h}_6^\dagger
                - i \hat{h}_6 \partial_\mu \hat{h}_6^\dagger    \\
            V_{6\mu}^\prime &= \hat{h}_6 V_{6\mu} \hat{h}_6^\dagger
                + \frac{1}{G_6} \hat{h}_6 \partial_\mu \hat{h}_6^\dagger.
       \end{array}
   \right. ,
\eea
\bea 
   \left\{
       \begin{array}{@{\,}ll} 
            \alpha_{\omega6\parallel\mu}^\prime &= 
                  \alpha_{\omega6\parallel\mu} 
                - i h_{Q1} \partial_\mu h_{Q1}^\dagger   \\
            V_{\omega6\mu}^\prime &= 
                  V_{\omega6\mu} 
              + \frac{1}{G_\omega6} h_{Q1}
                   \partial_\mu h_{Q1}^\dagger.  
       \end{array}
   \right. , 
\eea
\bea
   \left\{
       \begin{array}{@{\,}ll} 
            \alpha_{2\parallel\mu}^\prime &= 
                  \alpha_{2\parallel\mu} 
                - i \hat{h}_2 \partial_\mu \hat{h}_2^\dagger   \\
            V_{2\mu}^\prime &= 
                 V_{2\mu} 
                + \frac{1}{G_\omega2} \hat{h}_2 
                              \partial_\mu \hat{h}_2^\dagger . 
       \end{array}
   \right. , 
\eea
\bea
   \left\{
       \begin{array}{@{\,}ll} 
            \alpha_{\omega2\parallel\mu}^\prime &= 
                   \alpha_{\omega2\parallel\mu} 
                - i h_{L1} \partial_\mu h_{L1}^\dagger  \\
            V_{\omega2\mu}^\prime &= 
                 V_{\omega2\mu} 
                + \frac{1}{G_\omega2} h_{L1} \partial_\mu h_{L1}^\dagger.
       \end{array}
   \right. .  
\eea

In (\ref{eq:lv}), the terms which are proportional to $\alpha_V$
and $\beta_V$ break isospin symmetry. The term with the coefficient 
$\alpha_V$ generates mixing between $techni\ \omega_l$
($\omega_{2\mu}$  
($I=0$) ) and
$techni\ \rho_l$ ($\rho_{2\mu}$
($I=1$) )
through the kinetic term while the term which coefficient
is $\beta_V$ generates the mixing through the mass term.

Axial vector meson ($A$) and left-handed vector meson ($A_L$) sector\\
This part of the Lagrangian
are given by,
\bea
 {\cal L}_A &=& \frac{1}{2} tr F_{A_6}^{\mu\nu} {F_{A_6}}_{\mu\nu} 
            + \frac{1}{2} tr F_{A_L}^{\mu\nu} {F_{A_L}}_{\mu\nu}   
            + \frac{1}{2} tr F_{A_2}^{\mu\nu} {F_{A_2}}_{\mu\nu}   
            + \frac{1}{2} tr F_{A_{8}}^{\mu\nu} 
                                   {F_{A_{8}}}_{\mu\nu}
               \nn \\
         &-& M_{A_6}^2 
                  tr( A_{6\mu} -
                       \frac{i}{\ld_6}\alpha_{6\perp\mu} )^2
          - M_{A_8}^2 
                  tr( A_{8\mu} - \frac{i}{\ld_8}\alpha_{8\perp\mu} )^2
                    \nn \\                                  
          &-& M_{A_L}^2 
                  tr( A_{L\mu} - \frac{i}{\ld_L}\alpha_{L\perp\mu} )^2
           - M_{A_2}^2 
                  tr( A_{2\mu} 
                     - \frac{i}{\ld_2}
                           \alpha_{2\perp\mu} )^2 \nn \\
          &-& \alpha_A \frac{2}{\sqrt{3}} 
                  tr F_{A_2}^{\mu\nu} \tau^3
                                 {F_{A_{8}}}_{\mu\nu} \nn \\
          &+& \beta_A \frac{2}{\sqrt{3}}
                  tr( A_{2\mu} - \frac{i(1+\delta)}
                                  {\ld_2}\alpha_{2\perp\mu} )
                \tau^3
                    ( A_{8\mu}  
                     - \frac{i(1+\delta^\prime) }
                                  {\ld_{8}}\alpha_{8\perp\mu} ) 
                     \nn \\
         &+& \frac{1}{4}\frac{\beta_A^2 M_{A2}^2 \delta^2}
                      {M_{A2}^2 M_{A8}^2 -
                            \frac{1}{4}\beta_A^2}
                            \frac{1}{\ld_2^2}
                           tr( \alpha_{2\perp\mu})^2 \nn \\
          &+& \frac{1}{4} \frac{\beta_A^2 M_{A8}^2 {\delta^\prime}^2}
                      {M_{A2}^2 M_{A8}^2 -
                            \frac{1}{4}\beta_A^2}
                            \frac{1}{\ld_8^2}
                           tr( \alpha_{8\perp\mu})^2 \nn \\
         &+& \frac{2}{\sqrt{3}}\frac{\delta
                      \delta^\prime \beta_A M_{A2}^2 
                                              M_{A8}^2 }
                      {M_{A2}^2 M_{A8}^2 -
                            \frac{1}{4}\beta_A^2}
                         \frac{1}{\ld_2 \ld_8}
                        tr( \alpha_{2\perp\mu} \tau^3 
                        \alpha_{8\perp\mu} ).
\label{eq:lva}
 \eea
The decomposition into component fields is given by,
\bea
  A_{6\mu} &=& i\sum_{\alpha=1}^{35} \pmatrix{
                                       T^{\alpha } & \ \cr
                                       \ & 0 
                                             }a_{6\mu}^{\alpha }, \\
  A_{8\mu} &=& i\frac{1}{4\sqrt{3}} \pmatrix{
                                    I_6 & \ \cr
                                    \ & -3 I_2 
                                             } {a_{8\mu}}, \\
  A_{L\mu} &=& i\sum_{a=1}^{2} \frac{1}{2}\pmatrix{
                                     0 & \ \cr
                                     \ & \tau^a 
                                             } a_{L\mu}^a ,\\
  A_{2\mu} &=& i\frac{1}{2}\pmatrix{
                              0 & \ \cr
                              \ & \tau^3 
                                      } a_{2\mu}. 
\eea
In (46), $\tau^a$$(a=1,2)$ is a projection into $SU(2)_L$.
They transform under G in the following way,
\bea
 A_{6\mu}^\prime &=& \hat{h}_6 A_{6\mu} \hat{h}_6^\dagger ,\\
 A_{8\mu}^\prime &=& A_{8\mu} ,\\
 A_{L\mu}^\prime &=& \hat{h}_2 A_{L\mu} \hat{h}_2^\dagger, \\
 A_{2\mu}^\prime &=& A_{2\mu} . 
\eea
Note that there are not inhomogeneous terms for the transformation
of axial ($A$) and left-handed ($A_L$)
vector mesons under G. Therefore the mixing 
terms between these vector mesons and PNGB  occur. 
This effect results in redefinition of the coefficients of $ O(p^2)$
terms of NGB and PNGB
sector ((\ref{eq:S})). In order to  avoid the 
redifinition of  the coefficients, we just need to 
add the appropriate $O(p^2)$ terms in axial and left-handed vector 
meson sector. The $O(p^2)$ terms of NGBs and
 PNGBs in (\ref{eq:lva}) are chosen so that the $O(p^2)$
terms 
in (\ref{eq:S}) will
not be altered after eliminating $A$ and $A_L$ with their equation 
of motion. (See appendix {\bf D} for the details of the procedure.)
Compared with vector meson sector, two additional terms  
with the coefficients 
$\delta$ and $\delta^\prime$ come in (\ref{eq:lva}).  Both of them are isospin
breaking terms and can contribute to S and U through mixing 
between axial vector mesons ($a_2$ and $a_8$) and PNGBs ($\Pi^3$ and
$\theta_8$).
 However, because we focus on the contribution due to vector, axial 
and left-handed vector mesons  only, we ignore the contribution
due to these terms and put $\delta$ and $\delta^\prime$ zero
in the following analysis.

\section{ S,T and U parameter }

The electroweak radiative correction parameter S, T and U
are defined in terms of the gauge boson self-energy:\\ 
\bea
   S &=& 16\pi \frac{d}{dq^2}[ {\delta \Pi}_{33}(q^2) -
             {\delta \Pi}_{3Q}(q^2) ] \mid_{q^2 = 0 }  \nn  \\
    &=& - 16\pi \frac{d}{dq^2}[ {\delta \Pi}_{3Y}(q^2)]
                                      \mid_{q^2 = 0 }, \\
 \alpha T &=& \frac{g^2 + {g^\prime }^2}{M_Z^2}[ {\delta \Pi}_{11}(0) -
                                 {\delta \Pi}_{33}(0) ] , \\
  U &=& 16\pi \frac{d}{dq^2}[ {\delta \Pi}_{11}(q^2) -
                   {\delta \Pi}_{33}(q^2) ] \mid_{q^2 = 0 } .
\eea

where the $\delta\Pi $ is contribution of beyond standard model.

 In order to compute technimesons contribution to $S$, $T$ and
$U$, we need to expand the effective Lagrangian  in terms of
their component fields explicitly.  
Because we only consider their
tree level contribution here, it is suffice to keep
technicolor singlet and color singlet technimesons in the
expansion.
 We also have done one more simplification. In princple, not only
vector, axial and lefthanded-vector mesons, PNGBs can contribute
to $S$ and $U$ even in the tree level when isospin is not exact symmetry.
 For example,  PNGB $\theta_8$ can mix with
$W_3$ and $B$ through the isospin breaking term which strength is given by
$\beta_\perp$ in (\ref{eq:S}). Since the mixing term is derivative coupling,
the contribution to S is proportional to
$\beta_\perp^2/M_{\theta8}^2$. Here $M_{\theta 8}$ is a mass
of $\theta_8$which comes from the explicit breaking of $U(1)_{8A}$.
Similarly, $\Pi^3$ can contribute to $S$ and $U$ through isospin breaking terms
whose coefficients are $\delta$ and $\delta^\prime$. 
Since we have not known the masses of PNGBs and how they mix with each other,
we just assumed that they are so  heavy or the mixing with gauge
 bosons are so small that their contribution to S and U are negligible.
This amounts to the following simplification in the effective Lagrangian.
\bea
\beta_{\perp}=\delta=\delta^\prime=0.
\eea
Under this assumption, in tree level, it is spin 1 technimesons that 
contribute
to $S$ and $U$.
 In the following, we compute $S$, $T$ and $U$
 in 
 techniquark sector  and
in technilepton
sector  respectively. The latter computation will tell 
us how differently custodial
symmetry breaking terms contribute to S compared with  techniquark sector 
where custodial symmetry is exact.\\
 
  The techniquark sector 

In this sector, color singlet and $SU(2)$ triplet part of
techni-vector mesons $V_6 (1^{-})$ and $A_6 (1^{+})$
contribute to S. Since $SU(2)$ singlet vector mesons
such as $V_{\omega6}$ and $A_8$ do not couple with $W_3$,
they will not contribute to $S$. The mixing terms between
the vector mesons and gauge bosons are given by :
\bea
 {\cal L}_{6int} &=& - M_{V_6}^2 
                       tr( V_{6\mu} -
                           \frac{i}{G_6}\alpha_{6\parallel\mu} )^2 
                      - M_{A_6}^2 
                       tr( A_{6\mu} 
                          - \frac{i}{\ld_6}\alpha_{6\perp\mu} )^2
                    \nn \\                                  
                  &=& 2 M_{V_6}^2 [ \frac{\rho_6}{2} 
                            - \frac{\sqrt{3}}{4G_6}
                                   ( gW_3 +g^\prime B )]^2
                    + 2 M_{A_6}^2 [ \frac{a_6}{2} 
                            + \frac{\sqrt{3}}{4\ld_6}
                                   ( gW_3 - g^\prime B )]^2 .
\eea
$\rho_6$($a_6$) are color singlet and isotriplet
vector(axial vector) mesons.
Therefore, these terms induce $W_3-(\rho_6,a_6)$ and $B-(\rho_6,a_6)$
mixing and contributes to S through the Feynman diagram (Fig.(1)).
We obtain S in techniquark sector. 
\bea
  igg^\prime S_q = -16 \pi \frac{d}{d q^2} 
                     [(i \sqrt{3}\frac{M_{V6}^2}{2G_6}g)
                      \frac{-i}{q^2 - M_{V6}^2 }
                      (ig^\prime \sqrt{3} 
                          \frac{M_{V6}^2}{2G_6})]
                                         \mid_{q^2=0}
                         \nn \\
                   - 16 \pi \frac{d}{d q^2} 
                     [(i \sqrt{3}\frac{M_{A6}^2}{2\ld_6}g)
                      \frac{-i}{q^2 - M_{A6}^2 }
                      ( - ig^\prime \sqrt{3} 
                          \frac{ M_{A6}^2}{2\ld_6})]
                                         \mid_{q^2=0}.
\eea
\beq
    S_q = 4 \pi [ \frac{3}{G_6^2} - \frac{3}{\ld_6^2} ].
\label{eq:S6}
\eeq

\bigskip
\medskip

  The technilepton sector\\
In this sector, there are isospin breaking terms,
\bea
 {\cal L^\prime} &=& \alpha_V 
                  tr F_{V_2}^{\mu\nu} \tau^3
                                 {F_{V_\omega2}}_{\mu\nu} \nn \\
                &-& \beta_V 
                  tr( V_2 - \frac{i}{G_2}\alpha_{2\parallel\mu} )
                  \tau^3
                  ( V_{\omega2}  
                     - \frac{i}{G_{\omega2}}\alpha_{\omega2\parallel\mu} ) 
               \nn \\ 
                &-& \alpha_A \frac{2}{\sqrt{3}} 
                  tr F_{A_2}^{\mu\nu} \tau^3
                                 {F_{A_{8}}}_{\mu\nu} \nn \\
                &+& \beta_A \frac{2}{\sqrt{3}}
                  tr( A_{2\mu} - \frac{i
}
                                  {\ld_2}\alpha_{2\perp\mu} )
                            \tau^3
                    ( A_{8\mu}  
                     - \frac{i
 }
                                  {\ld_{8}}\alpha_{8\perp\mu} ) .
\eea
 Because of these terms, isospin of vector mesons 
is not a conserved quantity and mixing between I=0 and I=1
vector mesons can occur. Then, $SU(2)$ singlet vector mesons
such as $techni\ \omega_l$ ($\omega_2$) can contribute to S through the mixing
terms.
Therefore , we need to diagonalize
mass terms and kinetic terms of vector mesons.
By expanding the effective Lagrangian, we obtain: 
\bea
  {\cal L}_V&=& - \frac{1}{4}\pmatrix{
                 \partial_{[\mu,} \rho_{2\nu]} & 
                 \partial_{[\mu,} \omega_{2\nu]}
                              }
                     \pmatrix{
                         1 & \alpha_V \cr
                        \alpha_V & 1
                             }
                      \pmatrix{
                 \partial_{[\mu,} \rho_{2\nu]} \cr
                 \partial_{[\mu,} \omega_{2\nu]}
                              }
           \nn \\
    & & + \frac{1}{2}
           \pmatrix{
                \rho_2 & \omega_2 
                   }
           \pmatrix{
                M_{V2}^2 & \frac{\beta_V}{2} \cr
             \frac{\beta_V}{2} & M_{V_{\omega2}}
                   }
           \pmatrix{
                \rho_2 \cr
                \omega_2
                   } \nn \\
   & & -\frac{1}{2}
                  \pmatrix{
                            \rho_2 & \omega_2 
                          }
                  \pmatrix{
                            M_{V2}^2 & \frac{\beta_V}{2} \cr
                            \frac{\beta_V}{2} & M_{V_{\omega2}}^2 
                           }
                  \pmatrix{
                             \frac {gW_3 + g^\prime B}{G_2} \cr
                             \frac {2 Y_{Ll} g^\prime B}{G_{\omega2}}
                           } , 
\eea
\bea
  {\cal L}_A&=& - \frac{1}{4}\pmatrix{
                 \partial_{[\mu,} a_{2\nu]} & 
                 \partial_{[\mu,} a_{8\nu]}
                              }
                     \pmatrix{
                         1 & \alpha_A \cr
                        \alpha_A & 1
                             }
                      \pmatrix{
                 \partial_{[\mu,} a_{2\nu]} \cr
                 \partial_{[\mu,} a_{8\nu]}
                              }
           \nn \\
    & & + \frac{1}{2}
           \pmatrix{
                a_2 & a_8 
                   }
           \pmatrix{
                M_{A2}^2 & \frac{\beta_A}{2} \cr
             \frac{\beta_A}{2} & M_{A_{8}}
                   }
           \pmatrix{
                a_2 \cr
                a_8
                   } \nn \\
   & & +\frac{1}{2}
                  \pmatrix{
                            a_2 & a_8 
                          }
                  \pmatrix{
                            M_{A2}^2 & \frac{\beta_A}{2} 
\cr
                            \frac{\beta_A}{2}
 & M_{A_{8}}^2 
                           }
                  \pmatrix{
                             \frac {gW_3 - g^\prime B}{\ld_2} \cr
                                          0
                           } , 
\eea
where
\bea
  \partial_{[\mu,}V_{\nu]} = \partial_\mu V_\nu -
          \partial_\nu V_\mu . \nn 
\eea

The kinetic and mass terms can be diagonalized by doing the 
following transformations successively.
\bea
   \pmatrix{ 
             \rho_2^m \cr
             \omega_2^m \cr
            }
      =U_{mV} Z_V U_{DV}
   \pmatrix{
             \rho_2 \cr
             \omega_2 \cr
            },
\eea
\bea
   \pmatrix{ 
             a_2^m \cr
             a_8^m \cr
            }
      =U_{mA} Z_A U_{DA}
   \pmatrix{
             a_2 \cr
             a_8 \cr
            },
\eea

where $U_D$,$Z$ and $U_m$ are defined by,
\bea
    U_{DV(A)} &=&
   \pmatrix{
             \frac{1}{\sqrt{2}} & - \frac{1}{\sqrt{2}} \cr
             \frac{1}{\sqrt{2}} &   \frac{1}{\sqrt{2}} \cr
           }, \\
    Z_{V(A)} &=& 
   \pmatrix{
          (1-\alpha_{V(A)})^\frac{1}{2} & 0 \cr
                       0           &
                            (1+\alpha_{V(A)})^\frac{1}{2} 
            }, \\
    U_{mV(A)} &=& \pmatrix{
                c_{V(A)} & - s_{V(A)} \cr
                s_{V(A)} &   c_{V(A)} 
               }.
\eea
$U_D$ is a $45$ degree rotation matrix to diagonalize
the kinetic terms.
$Z$ is a scale transformation to keep correct
normalization for the kinetic terms. Finally $U_m$ is a rotation matrix which 
diagonalizes the mass terms.  $U_m$ relates the mass matrices to their eigen
values in the following way.
\bea 
 & \hspace{-1.5cm} \pmatrix{
                 M_\rho^2 & 0 \cr
                     0    & M_\omega^2   
                  }
       =
                  U_{mV}  Z_V^{-1} U_{DV}
            \pmatrix{
                 M_{V2}^2 & \frac{\beta_V}{2} \cr
                 \frac{\beta_V}{2} & M_{V\omega2}
                    }
                 U_{DV}^{-1}Z_V^{-1} U_{mV}^{-1}
                      \nn  \\
\medskip
  & \hspace{-0.5cm}
      = U_{mV}
           \pmatrix{
                 \frac{1}{2(1-\alpha_V)}(M_{V2}^2 
                              + M_{V\omega2}^2 - \beta_V) &
                 \frac{1}{2\sqrt{(1-\alpha_V^2)}
                                  }(M_{V2}^2 
                              - M_{V\omega2}^2 ) \cr
                 \frac{1}{2\sqrt{(1-\alpha_V^2)}}(M_{V2}^2 
                              - M_{V\omega2}^2 )        &
                 \frac{1}{2(1+\alpha_V)}(M_{V2}^2 
                              + M_{V\omega2}^2 + \beta_V) 
                  }
            U_{mV}^{-1},
\label{eq:57}
\eea
\bea 
   & \hspace{-1.5cm} \pmatrix{
                 M_{a2}^2 & 0 \cr
                     0    & M_{a8}^2   
                  }
       =
                  U_{mA}  Z_A^{-1} U_{DA}
            \pmatrix{
                 M_{A2}^2 & \frac{\beta_A}{2} \cr
                 \frac{\beta_A}{2} & M_{A8}
                    }
                 U_{DA}^{-1}Z_A^{-1} U_{mA}^{-1}
                      \nn \\
 & \hspace{-0.5cm}  = U_{mA}
           \pmatrix{
                 \frac{1}{2(1-\alpha_A)}(M_{A2}^2 
                              + M_{A8}^2 - \beta_A) &
                 \frac{1}{2\sqrt{(1-\alpha_A^2)}
                                  }(M_{A2}^2 
                              - M_{A8}^2 ) \cr
                 \frac{1}{2\sqrt{(1-\alpha_A^2)}}(M_{A2}^2 
                              - M_{A8}^2 )        &
                 \frac{1}{2(1+\alpha_A)}(M_{A2}^2 
                              + M_{A8}^2 + \beta_A) 
                  }
            U_{mA}^{-1}.
\label{eq:58}
\eea
In (\ref{eq:57}) and (\ref{eq:58}), $M_\rho$ and $M_\omega$ 
($M_{a2}$ and $M_{a8}$)
are eigenvalues of vector (axial vector) mesons mass matrix.
With these transformations, the interaction term between gauge 
bosons and vector mesons are given by:
\bea
   {\cal L}_{int} &=&  -\frac{1}{2}
                  \pmatrix{
                            \rho_2^m & \omega_2^m 
                          }
                  \pmatrix{
                           M_\rho^2 & 0 \cr
                             0 & M_\omega^2
                          }
                         U_{mV} Z_V U_{DV}
                   \pmatrix{
                             \frac {gW_3 + g^\prime B}{G_2} \cr
                             \frac {2 Y_{Ll} g^\prime B}{G_{\omega2}}
                           }  \nn \\
            & & +\frac{1}{2}
                  \pmatrix{
                            a_2^m & a_8^m 
                          }
                  \pmatrix{
                           M_{a2}^2 & 0 \cr
                             0 & M_{a8}^2
                          }
                         U_{mA} Z_A U_{DA}
                   \pmatrix{
                             \frac {gW_3 - g^\prime B}{\ld_2} \cr
                                           0
                           },
\eea
where,
\bea
        & \pmatrix{
                           M_\rho^2 & 0 \cr
                             0 & M_\omega^2
                          }
                         U_{mV} Z_V U_{DV}
                \nn \\
                 & \hspace{-1cm} =\frac{1}{\sqrt{2}}
                       \pmatrix{
                            M_{\rho}^2
                                     [c_V(1-\alpha_V)^{\frac{1}{2}}
                                      -s_V(1+\alpha_V)^{\frac{1}{2}}]
                          & M_{\rho}^2
                                     [-c_V(1-\alpha_V)^{\frac{1}{2}}
                                      -s_V(1+\alpha_V)^{\frac{1}{2}}]
                                         \cr
                           M_{\omega}^2 
                                     [c_V(1+\alpha_V)^{\frac{1}{2}}
                                      +s_V(1-\alpha_V)^{\frac{1}{2}}]
                          & M_{\omega}^2 
                                     [c_V(1+\alpha_V)^{\frac{1}{2}}
                                      -s_V(1-\alpha_V)^{\frac{1}{2}}]
                           }, 
\eea
\bea
          &\pmatrix{
                           M_{a2}^2 & 0 \cr
                             0 & M_{a8}^2
                          }
                         U_{mA} Z_A U_{DA}
                \nn \\
                    &\hspace{-1cm} =\frac{1}{\sqrt{2}}
                         \pmatrix{
                            M_{a2}^2
                                     [c_A(1-\alpha_A)^{\frac{1}{2}}
                                      -s_A(1+\alpha_A)^{\frac{1}{2}}]
                          & M_{a2}^2
                                     [-c_A(1-\alpha_A)^{\frac{1}{2}}
                                      -s_A(1+\alpha_A)^{\frac{1}{2}}]
                                         \cr
                            M_{a8}^2
                                     [c_A(1+\alpha_A)^{\frac{1}{2}}
                                      +s_A(1-\alpha_A)^{\frac{1}{2}}]
                          & M_{a8}^2
                                     [c_A(1+\alpha_A)^{\frac{1}{2}}
                                      -s_A(1-\alpha_A)^{\frac{1}{2}}]
                           }.
\eea
By computing the Feynman diagram (Fig.2),
the sum of the contribution to $S$ due to technilepton sector
 and   techniquark sector ((\ref{eq:S6})) is, 
\bea
    S &=& 4 \pi [ \frac{3}{G_6^2} - \frac{3}{\ld_6^2} 
              + \frac{1}{G_2^2} - \frac{1}{\ld_2^2} 
               + Y_{Ll} \frac{2 \alpha_V}{G_2 G_{\omega2}} ].
\label{eq:SS}
\eea
The term which is proportional to $Y_{Ll}$ is due to isospin breaking.
The minimum  of S can be obtained for $\alpha_V=1$ because of $ Y_{Ll}=-1$.
Only $\omega_2^m$ which consists of 
$N\bar{N}$ component  contributes to S in that case, because $E\bar{E}$
component of vector meson decouples and ideal mixing is realized.
This confirms that the conjecture of the importance of
$\omega$ which consists
of $N\bar{N}$ in the same model\cite{apel}.
For $T$ parameter, we obtain the same expression as that
is given in ref.\cite{apel}.
For completeness, we give the explicit form here.
\beq
\ M_W^2 = \frac{1}{4} g^2 \{ 3F_6^2 + F_L^2 \},
\eeq
\beq
\ M_Z^2 = \frac{1}{4} (g^2 + {g^\prime }^2) \{ 3F_6^2 + F_2^2 \},
\eeq
\beq
 \alpha T = \rho - 1 = \frac{F_L^2 - F_2^2}{3F_6^2 + F_2^2}.
\label{eq:T}
\eeq
With the assumption:$F_L, F_2 \ll F_6 $, $T$ parameter can be
so small even if there is a splitting between
$F_L$ and $F_2$ as stated by the authors of \cite{apel}.
 Finally we compute $U$ parameter. $U$ parameter is 
zero if the isospin symmetry is exact.
Therefore the contribution to $U$ in the present model
comes only from technilepton sector.
The   left-handed vector meson $a_L$ contributes to 
${\delta \Pi}_{11}$ part of $U$ (Fig.(3)). This contribution is 
not cancelled by ${\delta \Pi}_{33}$ part which 
$I=1$ vector ($\rho_2$) and axial vector ($a_2$)
mesons contribute to.
\bea
 U = 4 \pi [ \frac{1}{G_2^2} + \frac{1}{\ld_2^2} - \frac{1}{\ld_L^2} 
                   ]. 
\eea
The sign and the value of $U$ depend on three coupling constants
; $G_2$, $\lambda_2$, and $\lambda_L$.

\section{ The range of parameters for negative S}
In this section, we explore the parameter region where S parameter
is negative since the present experiment fits favor negative
$S$ \cite{sexp}. If the future experiments constraint on $S$
is improved, we may do more complete analysis. Before studying the results of the model without exact
custodial symmetry, let us review the estimation of $S_q$(the contribution
of S in techniquark sector) under the
assumption that the techniquark sector is just QCD
scale up technicolor theory so that we can get a hint of the order
of S which we are arguing.
If underlying dynamics of Technicolor for techniquark sector 
is exactly the same as that of QCD ($N_{TC}=N_{C}=3$), we can estimate
the parameters of the techniquark sector in the effective Lagrangian
by simply scaling up the corresponding parameters.
 Because $S_q$ depends
only on dimensionless coupling constants $G_6$ and $\lambda_6$, we 
may use the same value for the correspoinding coupling constants
which are determined by low energy hadronic processes. We have determined
$G_6$ from $\rho \rightarrow \pi \pi$ decay and $\lambda_6$
from $a_1(1260) \rightarrow \pi \gamma$.  This leads to
${G_6}^2=31.5$ and ${\lambda_6}^2=106$.(See appendix E.)
\beq
  S_q = (0.40-0.12)\times 3 =0.28 \times 3
\eeq
0.40 comes from the contribution of $I=1$ vector meson while
0.12 comes from $I=1$ axial vector meson.
A factor of 3 comes from the fact that the techniquark sector
contains three SU(2) doublet corresponding to color degree of freedom.
This result is consistent with the result with
dispersion analysis 
\cite{pes} and the results given by chiral lagrangian
with vector resonances. \cite{eck}

Now let us turn to the present model.
Since we have not known the underlying dynamics of the present model,
we do not have  any guiding principle to determine the parameters of our model
without experimental information.
Therefore instead of trying to predict S in our model, we 
determine the allowed region of the parameters of the effective 
lagrangian by imposing the present experimentally allowed region for S.
Since we have many parameters, we further need to  limit ourselves 
into the small parameter 
space to draw some definite conclusions. Here we simply assume that
S is dominated by only vector mesons and the contribution of axial 
vector mesons can be ignored. As shown in (78), this is a
good approximation in case
of the QCD scale up technicolor model. 
However it is not clear if the
assumption still holds even in the present model.
 Nevertheless, let us proceed further.
Under the assumption of vector ($1^{--}$) dominance,
$S$ is given by, 
\bea
  S &=&4\pi [ \frac{3}{G_6^2} 
                + \frac{1}{G_2^2} - {\alpha_V} \frac{2}{G_2 G_{\omega2}}]\nn \\
    &\ge&4\pi [ \frac{3}{G_6^2} 
                + \frac{1}{G_2^2} - \frac{2}{G_2 G_{\omega2}}].
\label{eq:Smin}
\eea
Here we have substituted $Y_{Ll}=-1$ in (\ref{eq:SS}) and neglect axial vector
contribution.
In the second line of (\ref{eq:Smin}), the inequality holds because
 $\alpha_V$ can take its value
between $-1$  and $1$ and 
$\alpha_V=1$ makes S minimum. 
Note that the bound for $\alpha_V$ comes from the condition
for the positive semi-definitness of the kinetic terms of
vector mesons (65).
In the follwing we assume that $\alpha_V=1$ and impose the condition of
negative S.
The condition for
$S \le 0$ now leads a relation,
\beq
 \frac{G_6}{G_2}( \frac{2G_6}{G_{\omega2}} - \frac{G_6}{G_2} )
                 \ge
   3 .
\eeq
This region is shown in Fig.(4) in the parameter
 space ($G_6/G_2$ vs $G_6/G_{\omega2}$).
We can get the lower bound for one of the parameter, $G_6/G_{\omega2}$.
\bea
 \frac{G_6}{G_{\omega2}} \ge \sqrt{3}
\eea  
Hence 
 $G_{\omega2}$ must be smaller than
$G_6$ in ordet to make S negative. 
This means that the coupling strength between $techni\ \omega_l$
and gauge bosons are stronger than that between $techni\ \rho_q$
and gauge bosons. Note that the coupling strength between
gauge bosons and vector mesons is proportional to $ 1/G$.
Since the negative contribution to S is proportional to $Y_{Ll}$
and $Y_{Ll}$  part of hypercharge interaction couples with 
$techni\ \omega_l$, strong coupling between $techni\ \omega_l$ and 
hypercharge gauge boson $B$ is preferred to get negative S.
  
\section{Conclusions and Discussion}
\par
 In this paper, we have constructed an effective 
Lagrangian for a technicolor model without exact custodial
symmetry. By using the Lagrangian, we compute
tree level contribution to
$S$ and $U$ due to spin 1 technimesons.
 We have shown that in  a realistic one-family model, 
the $techni\ \rho_l$ and the $techni\ \omega_l$ mixing  
can contribute to $S$ parameter with negative sign.  
The most important
term in our effective Lagrangian is the mixing in the kinetic
term, $ trF_{\rho } \tau^3 F_{\omega}$ . $S$ is independent
of the coeffcient of the mass mixing term $\beta_V$. 
We also study the condition to make S minimum under the
vector meson ($1^{--}$) dominance.
We find that the vector meson consists of $N\bar N$ component 
must be dominant dynamical degree of freedom to  make S minimum. ($\alpha_V=1$
) This argument holds as far as
hypercharge ($Y_{Ll}$) for which isospin breaking is introduced 
is negative.  Thus  the mechanism for negative S presented in this
paper does not work for one doublet model with $Y_L=0$. This
conclusion is consistent with an analysis with a free
technifermion model \cite{apel}.
On the contrary to the present model, we may introduce a small 
isospin breaking for techniquark sector. In that case,
the corresponding parameter of isospin breaking term, $\alpha_V$
must be -1 to make $S$ minimum because hypercharge of techniquark
$Y_{Lq}$ is positive. The vector mesons consisits of $D \bar D$
component will play major role to make $S$ minimum in that case. 
We also note that exotic lefthanded  charged vector mesons are naturally
introduced in our framework. They contribute to $U$ due to the mixing with 
$W^{\pm}$.  U can be both negative and positive depending on the parameters. 
There are many things to be done in this direction.
The relation between our computation and that with a free
technifermion model must be clarified.
The origin of the isospin breaking must be studied. Also we need to
relate the parameters of the effective Lagrangian to more
fundamental interaction for example, by modeling technicolor
by Nambu Jona-Lasinio model \cite{ono}.

{\Large{\bf Acknowledgement}}\\
We would like to thank V.A. Miransky for informing us of 
\cite{apel} and thank T. Onogi, K. Hikasa
and M. Harada for discussions.

\newpage

\appendix
\begin{center}
{\Large {\bf Appendix}}
\end{center}
\def\see{\setcounter{equation}{0}}
\renewcommand{\theequation}{ \Alph{section}.\arabic{equation} }
\see
In this appendix, we provide some useful formulas needed to
derive the results given in the text.
\section{Effective Lagrangian}   
\bea
 {\cal L}_S &=& F_6^2 tr( \alpha_{6\perp })^2 
          + F_L^2 tr( \alpha_{L\perp })^2 
          + F_2^2 tr( \alpha_{2\perp })^2 
          + F_8^2 tr( \alpha_{8\perp })^2 \nn \\
          &+& \beta_{\perp} tr( \alpha_{8\perp } \tau^3 \alpha_{2\perp} ) ,\\
               \nn \\
 {\cal L}_V &=& \frac{1}{2} tr F_{V_6}^{\mu\nu} {F_{V_6}}_{\mu\nu} 
            + \frac{1}{2} tr F_{V_2}^{\mu\nu} {F_{V_2}}_{\mu\nu}   
            + \frac{1}{2} tr F_{V_{\omega6}}^{\mu\nu} 
                                   {F_{V_{\omega6}}}_{\mu\nu}
            + \frac{1}{2} tr F_{V_{\omega2}}^{\mu\nu} 
                                   {F_{V_{\omega2}}}_{\mu\nu} \nn \\
           &-& M_{V_6}^2 
                  tr( V_{6\mu} - \frac{i}{G_6}\alpha_{6\parallel\mu} )^2
            - M_{V_{\omega6}}^2 
                  tr( V_{\omega6\mu}  
                     -\frac{i}{G_{\omega6}}
                            \alpha_{\omega6\parallel\mu} )^2  \nn \\
          &-& M_{V_2}^2 
                   tr( V_{2\mu} - \frac{i}{G_2}\alpha_{2\parallel\mu} )^2
            - M_{V_{\omega2}}^2 
                  tr( V_{\omega2\mu}  
                     - \frac{i}{G_{\omega2}}
                           \alpha_{\omega2\parallel\mu} )^2 \nn \\
          &+& \alpha_V 
                  tr F_{V_2}^{\mu\nu} \tau^3
                                 {F_{V_{\omega2}}}_{\mu\nu} \nn \\
          &-& \beta_V 
                  tr( V_{2\mu} - \frac{i}{G_2}\alpha_{2\parallel\mu} )
                \tau^3
                    ( V_{\omega2\mu}  
                     -\frac{i}{G_{\omega2}}\alpha_{\omega2\parallel\mu} ) ,\\
                     \nn   \\
 {\cal L}_A &=& \frac{1}{2} tr F_{A_6}^{\mu\nu} {F_{A_6}}_{\mu\nu} 
            + \frac{1}{2} tr F_{A_L}^{\mu\nu} {F_{A_L}}_{\mu\nu}   
            + \frac{1}{2} tr F_{A_2}^{\mu\nu} {F_{A_2}}_{\mu\nu}   
            + \frac{1}{2} tr F_{A_{8}}^{\mu\nu} 
                                   {F_{A_{8}}}_{\mu\nu}
               \nn \\
         &-& M_{A_6}^2 
                  tr( A_{6\mu} -
                       \frac{i}{\ld_6}\alpha_{6\perp\mu} )^2
          - M_{A_8}^2 
                  tr( A_{8\mu} - \frac{i}{\ld_8}\alpha_{8\perp\mu} )^2
                    \nn \\                                  
          &-& M_{A_L}^2 
                  tr( A_{L\mu} - \frac{i}{\ld_L}\alpha_{L\perp\mu} )^2
           - M_{A_2}^2 
                  tr( A_{2\mu} 
                     - \frac{i}{\ld_2}
                           \alpha_{2\perp\mu} )^2 \nn \\
          &-& \alpha_A \frac{2}{\sqrt{3}} 
                  tr F_{A_2}^{\mu\nu} \tau^3
                                 {F_{A_{8}}}_{\mu\nu} \nn \\
          &+& \beta_A \frac{2}{\sqrt{3}}
                  tr( A_{2\mu} - \frac{i(1+\delta)}
                                  {\ld_2}\alpha_{2\perp\mu} )
                \tau^3
                    ( A_{8\mu}  
                     - \frac{i(1+\delta^\prime) }
                                  {\ld_{8}}\alpha_{8\perp\mu} ) 
                     \nn \\
         &+& \frac{1}{4}\frac{\beta_A^2 M_{A2}^2 \delta^2}
                      {M_{A2}^2 M_{A8}^2 -
                            \frac{1}{4}\beta_A^2}
                            \frac{1}{\ld_2^2}
                           tr( \alpha_{2\perp\mu})^2 \nn \\
          &+& \frac{1}{4} \frac{\beta_A^2 M_{A8}^2 {\delta^\prime}^2}
                      {M_{A2}^2 M_{A8}^2 -
                            \frac{1}{4}\beta_A^2}
                            \frac{1}{\ld_8^2} 
                           tr( \alpha_{8\perp\mu})^2 \nn \\
         &+& \frac{2}{\sqrt{3}}\frac{\delta
                      \delta^\prime \beta_A M_{A2}^2
                                      M_{A8}^2 }
                      {M_{A2}^2 M_{A8}^2 -
                            \frac{1}{4}\beta_A^2}\frac{1}{\ld_2 \ld_8}
                        tr( \alpha_{2\perp\mu} \tau^3 
                        \alpha_{8\perp\mu} ).
 \eea
\see
\def\wa{( g W_\mu^3 - g^\prime B_\mu )}
\def\wv{( g W_\mu^3 + g^\prime B_\mu )}
\def\half{\frac{1}{2}}
\def\PI{\Pi}
\def\ti{\tilde}
\def\di{\partial_\mu}

\section{Decomposition into fields' components}
In terms of fields' components, $\alpha_{\perp}$s and
$\alpha_{\parallel}$s are given by:
\bea
\alpha_{6\perp\mu} &=& - \frac{1}{2\sqrt{3}F_6}
                  \{ \sum_{a=1}^2
                    \tau^a  ( \di P^a + \frac{\sqrt{3} F_6}{2} g W^a_\mu)   
                 \nn \\ 
       & &  \hspace{2cm}+ \tau^3 ( \di P^3 
                                 + \frac{\sqrt{3} F_6}{2} \wa )  
                  \} \ox I_3, \\
\alpha_{L\perp\mu} &=& - \frac{1}{2F_L} \sum_{a=1}^2 \tau^a
                  \{ \di\hat{\Pi}^a + \frac{F_L}{2} g W_\mu^a \} , \ \ 
                \hat{\Pi}^a \equiv \half \Pi^a ,\\
\alpha_{2\perp\mu} &=& - \frac{1}{2F_2} \tau^3
                   \{ \di\Pi^3 + \frac{F_2}{2} \wa \}, \\
\alpha_{8\perp\mu} &=& - \frac{1}{4\sqrt{3}F_8}
                       \pmatrix{
                          I_6 & \    \cr
                          \     & -3 I_2
                                }
                          \di\theta_8, \\
\eea
\bea
    A_{6\mu} - \frac{i}{\ld_6}\alpha_{6\perp\mu} 
        &=& \frac{i}{2\sqrt{3}} \tau^3 \ox I_3 [ a_{6\mu}^{3} 
                     + \frac{1}{\ld_6F_6}
                       \{ \di P^3 + \frac{\sqrt{3} F_6}{2} 
                        ( g W_\mu^3 - g^\prime B_\mu )
                       \}  
              \nn \\
        &+& \frac{i}{2\sqrt{3}} \sum_{a=1}^2 \tau^a \ox I_3
                     [ a_{6\mu}^{a} + \frac{1}{\ld_6 F_6}
                        \{ \di P^a + \frac{\sqrt{3} F_6}{2}
                                          gW^{a}_\mu  \} ],\\
  A_{L\mu} - \frac{i}{\ld_L}\alpha_{L\perp\mu} 
        &=& \frac{i}{2}\sum_{a=1}^2 \tau^a
                  [ a_{L\mu} + \frac{1}{\ld_LF_L}
                       \{ \di\hat{\Pi}^a + \frac{F_L}{2} 
                          gW_\mu^a \}],
\\
 A_{2\mu} - \frac{i}{\ld_2}\alpha_{2\perp\mu} 
        &=& \frac{i}{2} \tau^3 [ a_{2\mu} + \frac{1}{\ld_2 F_6}
                      \{ \di\Pi^3 
                     + \frac{F_2}{2}( gW_{3\mu} - g^\prime B_\mu ) \}],\\
 A_{8\mu} -
          \frac{i}{\ld_{8}}\alpha_{8\perp\mu} 
        &=&  \frac{i}{4\sqrt{3}}
               \pmatrix{
                       I_6 & \ \cr
                       \   & -3 I_2 
                       }
                    [ a_{8\mu} + \frac{1}{\ld_8 F_8} \di\theta_8 ], \\
   V_{6\mu} - \frac{i}{G_6}\alpha_{6\parallel\mu} 
        &=& \frac{i}{2\sqrt{3}} \tau^3 \ox I_3 
             [ \rho_{6\mu}^{3} - \frac{\sqrt{3}}{2G_6}
                          ( gW_{3\mu} + g^\prime B_\mu ) ]
              \nn \\
        & & - \frac{i}{2\sqrt{3}} \sum_{a=1}^2 \tau^a \ox I_3
             [ \rho_{6\mu}^{a} - \frac{\sqrt{3}}{2G_6}
                          ( gW_{\mu}^a ) ],   \\ 
   V_{\omega6\mu} -
          \frac{i}{G_{\omega6}}\alpha_{\omega6\parallel \mu}  
        &=& \frac{i}{2\sqrt{3}} I_6
             [ \omega_{6\mu} - \frac{\sqrt{3}}{2G_{\omega6}}
                          ( 2Y_{Ll} g^\prime B_\mu ) ], \\ 
   V_{2\mu} - \frac{i}{G_2}\alpha_{2\parallel\mu} 
        &=& \frac{i}{2} \tau^3 
             [ \rho_{2\mu} - \frac{1}{2G_2}
                          ( gW_{3\mu} + g^\prime B_\mu ) ],
                                             \\ 
   V_{\omega2\mu} -
           \frac{i}{G_{\omega2}}\alpha_{\omega2\parallel \mu}  
        &=& \frac{i}{2} I_2[ \omega_{2\mu} - \frac{1}{2G_{\omega2}}
                          ( 2Y_{Ll} g^\prime B_\mu ) ] . 
\eea

By substituting these expressions, we obtain :
\bea
{\cal L}_S =& &\half \{ (\di P^3)^2 + (\di\Pi^3)^2 +
                    (\di\theta_8)^2 \} 
            + \half \sum_{a=1}^2 \{ (\di P^a)^2 +
                            (\di\hat{\Pi}^a)^2 \} \nn \\
            &+& \half \sum_{a=1}^2
                            ( F_L \di\hat{\Pi}^a 
                              + \sqrt{3} F_6 \di P^a ) g W_\mu^a
            + \half ( F_2 \di\Pi^3 + \sqrt{3} F_6 \di P^3 )
                            \wa \nn \\
            &+& \frac{1}{8}(3F_6^2 + F_2^2)\wa^2
            + \frac{1}{8}(3F_6^2 + F_L^2) (g W_\mu^a)^2
                           \nn \\
            &-& \frac{\sqrt{3}\beta_\perp}{4F_8F_2}
                   \di\theta_8 \{ \di\Pi^3 + \frac{F_2}{2}
                                  \wa \} 
            + {\cal L}_{br}^\prime , \label{eq:A13} \\
                 \nn \\
 {\cal L}_V =&-&\frac{1}{4}(\partial_{[\mu,} \rho_{6\nu]})^2
              - \frac{1}{4}(\partial_{[\mu,} \omega_{6\nu]})^2
                 \nn \\
             &-&\frac{1}{4}(\partial_{[\mu,} \rho_{2\nu]})^2
              - \frac{1}{4}(\partial_{[\mu,} \omega_{2\nu]})^2        
              - \frac{\alpha_V}{2}(\partial_{[\mu,} \rho_{2\nu]})
                           (\partial_{[\mu,} \omega_{2\nu]})
                 \nn \\
             &+&\half M_{V6}^2\{ \rho_{6\mu}^3 -
                 \frac{\sqrt{3}}{2G_6} \wv \}^2
             +\half M_{V6}^2\{ \rho_{6\mu}^a 
                - \frac{\sqrt{3}}{2G_6} gW_\mu^a \}^2 \nn \\ 
             &+&\half M_{\omega6}^2\{ \omega_{6\mu} 
                - \frac{\sqrt{3}}{2G_{\omega6}} 2g^\prime
                         Y_{Lq}B_{\mu} \}^2 \nn \\
             &+&\half M_{V2}^2\{ \rho_{2\mu} 
                - \frac{1}{2G_2} \wv \}^2
              + \half M_{\omega2}^2 \{ \omega_{2\mu} 
                - \frac{1}{2G_{\omega2}} 2 g^\prime Y_{Ll}
                           B_\mu \}^2 \nn \\
             &+&\frac{\beta_V}{2} \{ \rho_{2\mu} - \frac{1}{2G_2}
                   \wv \}\{ \omega_{2\mu} 
                   - \frac{1}{2G_{\omega2}} 2 g^\prime
                                   Y_{Ll}B_\mu \},  \\
                  \nn \\
  {\cal L}_A =&-&\frac{1}{4}( \partial_{[\mu,} a_{6\nu]}^a )^2
               - \frac{1}{4}( \partial_{[\mu,} a_{8\nu]} )^2
                    \nn \\
              &-&\frac{1}{4}( \partial_{[\mu,} a_{2\nu]} )^2
               - \frac{\alpha_A}{2}
                 ( \partial_{[\mu,} a_{8\nu]} )
                 ( \partial_{[\mu,} a_{2\nu]} ) \nn \\
              &-&\frac{1}{4}\sum_{a=1}^2( \partial_{[\mu} a_{L\nu]}^a )^2 
                    \nn \\
              &+&\half M_{A6}^2 \sum_{a=1}^2[ a_{6\mu}^a +
                 \frac{1}{\ld_6 F_6 } \{ \di P^a +
                  \frac{\sqrt{3}F_6}{2} gW_\mu^a \}]^2 
                  \nn \\
              &+&\half M_{A6}^2 [ a_{6\mu}^3 
                 + \frac{1}{\ld_6 F_6}\{ \di P^3 +
                   \frac{\sqrt{3}F_6}{2}\wa \} ]^2
                  \nn \\
              &+&\half M_{A8}^2 [ a_{8\mu} 
                 + \frac{1}{\ld_8 F_8}\di\theta_8 ]^2 \nn \\
              &+&\half M_{A2}^2 [ a_{2\mu}^3
                 + \frac{1}{\ld_2 F_2} \{ \di\Pi^3 
                 + \frac{F_2}{2}\wa \} ]^2 \nn \\
              &+&\frac{\beta_A}{2} [ a_{2\mu}^3 
                 + \frac{1+\delta}{\ld_2 F_2} \{ \di\Pi^3
                 + \frac{F_2}{2}\wa \} ]
                   [ a_{8\mu} 
                 + \frac{1+\delta^\prime}{\ld_8 F_8}\di\theta_8 ] \nn \\
              &+&\half M_{AL}^2 \sum_{a=1}^2[ a_{L\mu}^a 
                 + \frac{1}{\ld_L F_L}\{ \di\hat{\Pi}^a 
                 + \frac{F_L}{2} g W_\mu^a \} ]^2 \nn \\
          &+& \frac{1}{8}\frac{\beta_A^2 M_{A2}^2 \delta^2}
                      {M_{A2}^2 M_{A8}^2 -
                            \frac{1}{4}\beta_A^2}
                           \frac{1}{\ld_2^2 F_2^2}
                           \{ \di \Pi^3 + \frac{F_2}{2} \wa \}^2 \nn \\
          &+& \frac{1}{8} \frac{\beta_A^2 M_{A8}^2 {\delta^\prime}^2}
                      {M_{A2}^2 M_{A8}^2 -
                            \frac{1}{4}\beta_A^2}
                          \frac{1}{\ld_8^2 F_8^2}
                          \di \theta_8^2 \nn \\
         &-& \frac{1}{2}\frac{\delta
                      \delta^\prime \beta_A M_{A2}^2 M_{A8}^2}
                      {M_{A2}^2 M_{A8}^2 -
                            \frac{1}{4}\beta_A^2}\frac{1}{F_2 F_8 \ld_2 \ld_8}
                        \di \theta_8\{ \di \PI^3 
                         + \frac{F_2}{2} \wa \},
\label{eq:A15}
\eea
We fix the gauge into unitary gauge. This corresponds to
the follwing replacement. 
\def\ra{\rightarrow}
\bea
  \wa &\ra& \wa - \frac{2}{3F_6^2 + F_2^2}
                   (\sqrt{3}F_6 \di P^3 + F_2 \di \Pi^3 ),
                              \\
  gW_\mu^a &\ra& gW_\mu^a - \frac{2}{3F_6^2 + F_L^2}
                   (\sqrt{3}F_6 \di P^a + F_2 \di
                                              \hat{\Pi}^a ),
\eea
We also redefine the axial vector and left-handed vector mesons.
\bea
  a_{6\mu}^a \ &\ra& \ a_{6\mu}^a - \frac{1}{\ld_6 F_6} \di P^a,
\\
  a_{6\mu}^3 \ &\ra& \ a_{6\mu}^3 - \frac{1}{\ld_6 F_6} \di
                                                       P^3, \\
  a_{8\mu} \ &\ra& \ a_{8\mu} - \frac{1}{\ld_8 F_8} \di
                                                   \theta_8, \\
  a_{2\mu}^3 \ &\ra& \ a_{2\mu}^3 - \frac{1}{\ld_2 F_2} \di
                                                 \Pi^3, \\
  a_{L\mu}^a \ &\ra& \ a_{L\mu}^a - \frac{1}{\ld_L F_L} 
                                 \di \hat{\Pi}^a.
\eea
Then, ${\cal L}_S$ and ${\cal L}_A$ are written in terms of
physical degrees of freedom. 
\bea
{\cal L}_S &=& \frac{1}{2} \{ (\di \ti{\Pi}^3 )^2 
                           +(\di \ti{\Pi}^a)^2
                           +(\di \theta_8)^2 \} \nn \\
         & & - \frac{\sqrt{3} \beta_\perp }{4 F_8 F_2}
                         \partial \theta_8
                      \{ \cos \chi_3 \di \ti{\Pi}^3
                           + \frac{F_2}{2} 
                                \wa \} \nn \\
         & & + \frac{1}{8}( 3F_6^2 + F_2^2 ) \wa^2
             + \frac{1}{8}( 3F_6^2 + F_L^2 ) (gW_\mu^a)^2
             \nn \\
         & & + {\cal L}_{br}^\prime ,\\
              \nn \\
  {\cal L}_A =&-&\frac{1}{4}( \partial_{[\mu,} a_{6\nu]}^a )^2
               - \frac{1}{4}( \partial_{[\mu,} a_{8\nu]} )^2
                    \nn \\
              &-&\frac{1}{4}( \partial_{[\mu,} a_{2\nu]} )^2
               - \frac{\alpha_A}{2}
                 ( \partial_{[\mu,} a_{8\nu]} )
                 ( \partial_{[\mu,} a_{2\nu]} ) \nn \\
              &-&\frac{1}{4}\sum_{a=1}^2( \partial_{[\mu,} a_{L\nu]}^a )^2 
                    \nn \\
              &+&\half M_{A6}^2 \sum_{a=1}^2[ a_{6\mu}^a +
                  \frac{\sqrt{3}}{2\ld_6} gW_{\mu}^a ]^2 
                  \nn \\
              &+&\half M_{A6}^2 [ a_{6\mu}^3 
                 +  
                   \frac{\sqrt{3}}{2\ld_6}\wa ]^2
                  \nn \\
              &+&\half M_{A8}^2 [ a_{8\mu}]^2 \nn \\
              &+&\half M_{A2}^2 [ a_{2\mu}
                 + \frac{1}{2\ld_2}  
                  \wa  ]^2 \nn \\
              &+&\frac{\beta_A}{2} [ a_{2\mu} 
                 + \frac{1}{2\ld_2} 
                     \wa ]
                   [ a_{8\mu} 
                 ] \nn \\
              &+&\half M_{AL}^2 \sum_{a=1}^2[ a_{L\mu}^a 
                 + \frac{1}{2\ld_L}  
                    g W_\mu^a ]^2 \nn \\
             &+&\frac{\beta_A \delta^\prime}{2 \ld_8 F_8}
                   a_2^3 \di \theta_8 
             + \frac{\beta_A \delta}{2\ld_2 F_2}
                   a_8 \{ cos\chi_3 \di \ti{\Pi}^3 +
                  \frac{F_2}{2}\wa \} \nn \\
          &+& \frac{1}{8}\frac{\beta_A^2 M_{A2}^2 \delta^2}
                      {M_{A2}^2 M_{A8}^2 -
                            \frac{1}{4}\beta_A^2}
                           \frac{1}{\ld_2^2 F_2^2}
                           \{ cos \chi_3 \di \ti{\Pi}^3 
                              + \frac{F_2}{2} \wa \}^2 \nn \\
          &+& \frac{1}{8} \frac{\beta_A^2 M_{A8}^2 {\delta^\prime}^2}
                      {M_{A2}^2 M_{A8}^2 -
                            \frac{1}{4}\beta_A^2}
                          \frac{1}{\ld_8^2 F_8^2}
                          (\di \theta_8)^2  \\
         &-& \frac{1}{8}\frac{\delta
                      \delta^\prime \beta_A^3 }
                      {M_{A2}^2 M_{A8}^2 -
                            \frac{1}{4}\beta_A^2}\frac{1}{F_2 F_8 \ld_2 \ld_8}
                      \di \theta_8\{ cos\chi_3 \di \ti{\PI}^3 
                         + \frac{F_2}{2} \wa \},\nn
\eea
where,
\bea
\pmatrix{
         \ti{\Pi}^a \cr
         \ti{P}^a
         }
          &=&
         \pmatrix{
                cos\chi_L & -sin\chi_L \cr
                sin\chi_L &  cos\chi_L
                 }
         \pmatrix{
                 \hat{\Pi}^a \cr
                 P^a
                  },  \\
\pmatrix{
         \ti{\Pi}^3 \cr
         \ti{P}^3
         }
          &=&
         \pmatrix{
                cos\chi_3 & -sin\chi_3 \cr
                sin\chi_3 &  cos\chi_3
                   }
         \pmatrix{
                \Pi^3 \cr
                P^3
                  } ,
\eea
\bea
   & & cos\chi_L = \frac{\sqrt{3}F_6}{\sqrt{3F_6^2 + F_L^2}} ,
   \ \ \  sin\chi_L = \frac{F_L}{\sqrt{3F_6^2 + F_L^2}}, \\
   & & cos\chi_3 = \frac{\sqrt{3}F_6}{\sqrt{3F_6^2 + F_2^2}} ,
   \ \ \   sin\chi_3 = \frac{F_2}{\sqrt{3F_6^2 + F_2^2}}. 
\eea

\see
\section{Pseudo Nambu Goldstone boson sector}
\def\PI{\Pi}
\def\ti{\tilde}
\bea
   {\cal L}_S &=& \frac{1}{2} \{ (\di \ti{\Pi}^3 )^2 
                           +(\di \ti{\Pi}^a)^2
                           +(\di \theta_8)^2 \} \nn \\
            &+& \frac{1}{8}(3F_6^2 + F_2^2)\wa^2
            + \frac{1}{8}(3F_6^2 + F_L^2) (g W_\mu^a)^2
                           \nn \\
         & & - \frac{\sqrt{3} \beta_\perp }{4 F_8 F_2}
                         \partial \theta_8
                      \{ \cos \chi_3 \di \ti{\Pi}^3
                           + \frac{1}{2} 
                                  \wa \}
            + {\cal L}_{br}^\prime.
\eea
Now we are ready for giving the explicit form for ${\cal L}_{br}^\prime$.
Because $\ti{\Pi}^3$, $\ti{\Pi}^a$ and $\theta_8$ are NGBs
associated with broken global symmetry, we can introduce the
follwing  mass terms as ${\cal
L}_{br}^\prime$.
\bea
 {\cal L}_{br}^\prime = - \frac{1}{2}
                       \pmatrix{
                           \ti{\Pi}_3 & \theta_8
                               }
                       \pmatrix{
                            m_3^2 & \frac{\beta_S}{2} \cr
                            \frac{\beta_S}{2}  & m_8^2 
                               }
                       \pmatrix{
                            \ti{\Pi}_3 \cr
                            \theta_8
                                }
              - \frac{1}{2} m_a^2 \ti{\Pi}_a^2
\eea
The mass terms  break the global
symmetry without loss of $SU(2) \ox  U(1)$ gauge invariance 
and these NGBs become  PNGBs.

\see
\section{$O(p^2)$ terms in axial vector and left-handed vector sector}
 In this appendix, we show how to determine the $O(p^2)$
terms which consist of NGBs and PNGBs in this sector.
As explained in the text, we add the $O(p^2)$ terms so that the
incorporation of axial and left-handed vector mesons does not 
change the decay constants of NGBs and PNGBs. By doing so, $\rho$ $(T)$
parameter  depends only on the parameters in ${\cal L}_s$
.  
  Though it is just the matter of the definition
of the parameters of $O(p^2)$ terms in ${\cal L}_s$, our choise is convenient
because the the parameters in ${\cal L}_s$ are directly related to  
physical quantities such as decay constants. Further $\rho$
$(T)$ parameter is independent of the parameters in ${\cal
L}_A$ with the procedure adopted here.
Let us discuss $0(p^2)$ terms which consist of
$\alpha_{6\perp}$ and 
$\alpha_{L\perp}$ first. These terms are related to $A_6$ and $A_L$.
The equations of motion of $A_6$ and $A_L$
  up to $O(p^2)$ are,
\bea
A_6 &=&\frac{i}{\ld_6}\alpha_{6\perp},  \nn \\
A_L &=&\frac{i}{\ld_L}\alpha_{L\perp}. 
\eea
By substituting these into (44), we do not have $O(p^2)$ terms of NGBs 
and PNGBs. Therefore the $O(p^2)$ terms which are  already present 
in (44) is enough.
$O(p^2)$ terms which consist of $\alpha_{2\perp}$ and $\alpha_{8\perp}$
have more complicated  coefficients as shown in (44).
We have determined them in the follwing way.
Let us focus on a part of ${\cal L}_A$
\bea
     {\cal L}_0
          =& & \frac{1}{2} M_{A_8}^2 
                  (a_{8\mu} - \frac{1}{\ld_8}\hat{\alpha}_{8\perp\mu} )^2
             + \frac{1}{2} M_{A_2}^2 
                  ( a_{2\mu} 
                     - \frac{1}{\ld_2}
                          \hat{ \alpha}_{2\perp\mu} )^2 \nn \\
          &+& \beta_A \frac{1}{2}
                  ( a_{2\mu} - \frac{(1+\delta)}
                                  {\ld_2}\hat{\alpha}_{2\perp\mu} )
                    ( a_{8\mu}  
                     - \frac{(1+\delta^\prime) }
                                  {\ld_{8}}\hat{\alpha}_{8\perp\mu} ) \nn \\
                \nn \\
          =& &\frac{1}{2} \pmatrix{
                    (a_2 -
                   \frac{1}{\ld_2}\hat{\alpha}_{2\perp } ) &
                    (a_8 -
                   \frac{1}{\ld_8}\hat{\alpha}_{8\perp } )
                                  }
                          \pmatrix{
                     M_{A2}^2 & \frac{\beta_A}{2} \cr
                     \frac{\beta_A}{2} & M_{A8}^2 
                                  }
                           \pmatrix{
                     (a_2 -
                   \frac{1}{\ld_2}\hat{\alpha}_{2\perp } ) \cr
                    (a_8 -
                   \frac{1}{\ld_8}\hat{\alpha}_{8\perp } )
                                  } \nn \\
       & &+ \frac{1}{2} \pmatrix{
                     (a_2 -
                   \frac{1}{\ld_2}\hat{\alpha}_{2\perp } ) &
                    (a_8 -
                   \frac{1}{\ld_8}\hat{\alpha}_{8\perp } )
                                  }
                        \pmatrix{
                      0 & \beta_A \delta^\prime \cr
                      \beta_A \delta & 0
                                  }
                         \pmatrix{
                  - \frac{1}{\ld_2}\hat{\alpha}_{2\perp }  \cr
                  - \frac{1}{\ld_8}\hat{\alpha}_{8\perp } 
                                  } \nn \\
           & & + \frac{\beta_A}{2}
                   \frac{\delta\delta^\prime}{\ld_2 \ld_8}
                  \hat{\alpha}_{2\perp } \hat{\alpha}_{8\perp},   
\label{eq:lax}                       
\eea
where we have used the follwing notation;
\bea
 A_{2\mu} &=& i \frac{1}{2}\pmatrix{
                          0 & \ \cr
                          \ & \tau^3
                                   } a_{2\mu}, \\
 A_{8\mu} &=& i \frac{1}{4\sqrt{3}}\pmatrix{
                    I_6 & \ \cr
                    \ & -3 I_2
                         }
                      a_{8\mu} ,
\eea
\bea
 \alpha_{2\perp\mu} &=&
\frac{1}{2}\pmatrix{
                  0 & \ \cr
                 \ & \tau^3
                    }\hat{\alpha}_{2\perp\mu} \\
 \alpha_{8\perp\mu} &=&
         \frac{1}{4\sqrt{3}}\pmatrix{
                             I_6 & \ \cr
                             \  & -3 I_2
                                      }
               \hat{\alpha}_{8\perp\mu}
\eea
The equations of motion for $a_2$ and $a_8$ up to $O(p^2)$ are:
\bea
              \pmatrix{
                    a_2 -
                   \frac{1}{\ld_2}\hat{\alpha}_{2\perp }  \cr
                    a_8 -
                   \frac{1}{\ld_8}\hat{\alpha}_{8\perp } 
                                  } 
        = \frac{1}{2}
          \frac{1}{M_{A2}^2 M_{A8}^2 - \frac{\beta_A^2}{4}}
              \pmatrix{
                   M_{A8}^2 & - \frac{\beta_A}{2} \cr
                 - \frac{\beta_A}{2} & M_{A2}^2 
                      }
              \pmatrix{
                      0 & \beta_A \delta^\prime \cr
                    \beta_A \delta & 0
                      }
              \pmatrix{
                   \frac{1}{\ld_2}\hat{\alpha}_{2\perp }  \cr
                   \frac{1}{\ld_8}\hat{\alpha}_{8\perp } 
                                  } 
\eea
By substituting this into ${\cal L}_0$,
we obtain;
\bea 
 {\cal L}_c =    & & -\frac{1}{4}\frac{\beta_A^2 M_{A2}^2 \delta^2}
                      {M_{A2}^2 M_{A8}^2 -
                            \frac{1}{4}\beta_A^2}
                            \frac{1}{\ld_2^2}
                           tr( \alpha_{2\perp\mu})^2 \nn \\
          &+& -\frac{1}{4} \frac{\beta_A^2 M_{A8}^2 {\delta^\prime}^2}
                      {M_{A2}^2 M_{A8}^2 -
                            \frac{1}{4}\beta_A^2}
                            \frac{1}{\ld_8^2}
                           tr( \alpha_{8\perp\mu})^2 \nn \\
         &+& -\frac{2}{\sqrt{3}}\frac{\delta
                      \delta^\prime \beta_A M_{A2}^2 
                                              M_{A8}^2 }
                      {M_{A2}^2 M_{A8}^2 -
                            \frac{1}{4}\beta_A^2}
                         \frac{1}{\ld_2 \ld_8}
                        tr( \alpha_{2\perp\mu} \tau^3 
                        \alpha_{8\perp\mu} )
\label{eq:D3}
\eea
By subtracting ${\cal L}_c$ from  ${\cal L}_0$, we obtain
$ O(p^2) $terms which consist of $\alpha_{2 \perp}$ and
$\alpha_{8 \perp}$ in (44). 
These are the desired counter terms 
which kill the effect of axial vector mesons and left-handed 
vector mesons on $\rho$ $(T)$ parameter.

\see
\section{ S in QCD scale up technicolor model}
For the completeness, we compute  $S_{theory}$ of
QCD scale up technicolor model 
which is quoted
in the introduction and (78).
The QCD scale up technicolor model 
has $SU(2)_L \ox SU(2)_R$ global symmetry and $N_{TC}=N_c=3$.
Therefore we only need to study the SU(2) subsector of one-family model.
$ S$ in this model is given by;
\beq
 S = 4\pi [ \frac{1}{G_V^2} - \frac{1}{\ld_A^2} ].
\eeq
The $G_V$ and $\ld_A$ are defined in the same way as $G_6$ and $\ld_6$.
$G_V$ and $\ld_A$ are determined by 
$\rho \rightarrow \pi\pi$ and $a_1 \rightarrow \gamma\pi$ decays
.
\bea
 \Gamma_{\rho \pi \pi} &=& \frac{1}{48\pi} m_{\rho} 
                  ( \frac{m_{\rho}^2}{2G_V f_{QCD}^2})^2 
                  ( 1 - \frac{4m_\pi^2}{m_{\rho}^2} )^{\frac{3}{2}},
          \\
 \Gamma_{a \gamma \pi} &=&  \frac{\alpha}{24 f_{QCD}^2 G_A^2}
                          \frac{(m_a^2 - m_\pi^2 )^3}{m_a^3},
\eea   
where $f_{QCD}$ is the pion decay constant.
By using the following values, 
\bea
 &f_{QCD} = 93 MeV,\ \ \ m_\pi = 140 MeV, \ \ \ m_\rho = 768
              MeV, \ \ \ m_a = 1260 MeV, \nn \\
 &\Gamma_{\rho \pi \pi} = 152 MeV , \ \ \ \ \Gamma_{a \gamma
                     \pi} = 0.64 MeV, \nn
\eea
we obtain;
\bea
 G_V^2 = 31.5,\ \ \ \ G_A^2 = 106 . \nn
\eea
This leads to the following estimation of  $S_{theory}$ for the
QCD scale up technicolor model which is quoted in the text.
\beq
 S_{theory} = 0.28 
\eeq
Because they are dimensionless quantities,
the scaling relations between the parameters in QCD and that of
the
technicolor are given by,
\bea
 G_6=G_V  ,\ \  \ld_6=\ld_A.
\eea
This equalities hold if the underlying dynamics of the techniquark sector is the same as that
of QCD. 

\newpage 

\newpage
\begin{center}
{\Large{\bf Table Caption}}
\end{center}
\begin{itemize}
\item {\bf Table(1)}: J,P,C,I and electric charge of the technimesons
incorporated in the effective Lagrangian. In the third column, 
the numbers of corresponding technimesons are shown. In the fourth
column, techniquark contents with the same J,P,C,I
and electric charge are shown.
\end{itemize}

\begin{center}
{\Large{\bf Figure Captions}}
\end{center}
\begin{itemize}
\item {\bf Figure(1)}: The Feynman diagrams for the
contribution
to
$S$ in the techniquark sector.

\item {\bf Figure(2)}: The Feynman daigrams for the
contribution to
 $S$ in the 
technilepton sector.

\item {\bf Figure(3)}: The Feynman diagram for the
contribution
to $\delta \Pi_{11}$ part of
$U$ due to the left-handed vector meson.

\item {\bf Figure(4)}: $S < 0 $ $(S > 0)$
region with vector( $1^{--}$) dominance
assumption for $\alpha_V = 1$. $G_6, G_2$ and $G_{\omega2}$
are coupling constants associated with vector mesons. (See
(\ref{eq:lv}) in the text.) 
\end{itemize}

\newpage

{\Large
\[ \begin{array}{|c|c|c|c|c|c|}  \hline
 & J^{pc} & Number &   & Isospin & charge \\ \hline
 P^\alpha & 0^- & 35 & \bar{Q} \gamma_5 T^\alpha Q &  &   \\ \hline
 \Pi^3 & 0^{-+} & 1 & \bar{L}\gamma_5 \tau^3 L & 1 & 0 \\ \hline
 \theta_8 & 0^{-+} & 1 & \bar{Q}\gamma_5 Q - 3 \bar{L}\gamma_5 L
                & 0 & 0 \\ \hline
 \Pi^\pm & 0 & 2 & \bar{L} \frac{1}{2}(1-\gamma_5
                                        )\tau^{\pm} L 
                & 1 & \pm 1 \\ \hline
 V_{6\mu}^\alpha & 1^- & 35 & \bar{Q} \gamma_\mu \ld^\alpha Q
                & &    \\ \hline
 \omega_{6\mu} & 1^{--} & 1 & \bar{Q} \gamma_\mu Q
                & 0 & 0 \\ \hline 
 \rho_{2\mu} & 1^{--} & 1 & \bar{L} \gamma_\mu \tau^3 L 
                & 1 & 0 \\ \hline
 \omega_{2\mu} & 1^{--} & 1 & \bar{L} \gamma_\mu L & 0 & 0 \\ \hline
 A_{6\mu}^\alpha & 1^+ & 35 & \bar{Q} \gamma_\mu \gamma_5
                                    T^\alpha Q 
                & &   \\ \hline
 A_{8\mu} & 1^{++} & 1 & \bar{Q} \gamma_\mu \gamma_5 Q - 3
                                 \bar{L}\gamma_\mu \gamma_5 L
                & 0 & 0  \\ \hline
 A_{2\mu} & 1^{++} & 1 & \bar{L} \gamma_\mu \gamma_5 \tau^3 L
                & 1 & 0  \\ \hline
 A_{L\mu}^\pm & 1 & 2 & \bar{L} \gamma_\mu \frac{1}{2} (1-\gamma_5) \tau^\pm
                                             L
                & 1 & \pm 1 \\ \hline
 \end{array}
\]}

\begin{center}
{\bf Table 1}
\end{center}

\newpage

\newread\epsffilein    
\newif\ifepsffileok    
\newif\ifepsfbbfound   
\newif\ifepsfverbose   
\newdimen\epsfxsize    
\newdimen\epsfysize    
\newdimen\epsftsize    
\newdimen\epsfrsize    
\newdimen\epsftmp      
\newdimen\pspoints     
\pspoints=1bp          
\epsfxsize=0pt         
\epsfysize=0pt         
\def\epsfbox#1{\global\def\epsfllx{72}\global\def\epsflly{72}%
   \global\def\epsfurx{540}\global\def\epsfury{720}%
   \def\lbracket{[}\def\testit{#1}\ifx\testit\lbracket
   \let\next=\epsfgetlitbb\else\let\next=\epsfnormal\fi\next{#1}}%
\def\epsfgetlitbb#1#2 #3 #4 #5]#6{\epsfgrab #2 #3 #4 #5 .\\%
   \epsfsetgraph{#6}}%
\def\epsfnormal#1{\epsfgetbb{#1}\epsfsetgraph{#1}}%
\def\epsfgetbb#1{%
%
%
\openin\epsffilein=#1
\ifeof\epsffilein\errmessage{I couldn't open #1, will ignore it}\else
%
%
   {\epsffileoktrue \chardef\other=12
    \def\do##1{\catcode`##1=\other}\dospecials \catcode`\ =10
    \loop
       \read\epsffilein to \epsffileline
       \ifeof\epsffilein\epsffileokfalse\else
%
%
          \expandafter\epsfaux\epsffileline:. \\%
       \fi
   \ifepsffileok\repeat
   \ifepsfbbfound\else
    \ifepsfverbose\message{No bounding box comment in #1; using defaults}\fi\fi
   }\closein\epsffilein\fi}%
%
%
\def\epsfsetgraph#1{%
   \epsfrsize=\epsfury\pspoints
   \advance\epsfrsize by-\epsflly\pspoints
   \epsftsize=\epsfurx\pspoints
   \advance\epsftsize by-\epsfllx\pspoints
%
%
   \epsfxsize\epsfsize\epsftsize\epsfrsize
   \ifnum\epsfxsize=0 \ifnum\epsfysize=0
      \epsfxsize=\epsftsize \epsfysize=\epsfrsize
%
%
     \else\epsftmp=\epsftsize \divide\epsftmp\epsfrsize
       \epsfxsize=\epsfysize \multiply\epsfxsize\epsftmp
       \multiply\epsftmp\epsfrsize \advance\epsftsize-\epsftmp
       \epsftmp=\epsfysize
       \loop \advance\epsftsize\epsftsize \divide\epsftmp 2
       \ifnum\epsftmp>0
          \ifnum\epsftsize<\epsfrsize\else
             \advance\epsftsize-\epsfrsize \advance\epsfxsize\epsftmp \fi
       \repeat
     \fi
   \else\epsftmp=\epsfrsize \divide\epsftmp\epsftsize
     \epsfysize=\epsfxsize \multiply\epsfysize\epsftmp
     \multiply\epsftmp\epsftsize \advance\epsfrsize-\epsftmp
     \epsftmp=\epsfxsize
     \loop \advance\epsfrsize\epsfrsize \divide\epsftmp 2
     \ifnum\epsftmp>0
        \ifnum\epsfrsize<\epsftsize\else
           \advance\epsfrsize-\epsftsize \advance\epsfysize\epsftmp \fi
     \repeat
   \fi
%
%
   \ifepsfverbose\message{#1: width=\the\epsfxsize, height=\the\epsfysize}\fi
   \epsftmp=10\epsfxsize \divide\epsftmp\pspoints
   \vbox to\epsfysize{\vfil\hbox to\epsfxsize{%
      \includegraphics{#1}%
      \hfil}}%
\epsfxsize=0pt\epsfysize=0pt}%
 
%
%
{\catcode`\%=12 \global\let\epsfpercent=
%
%
\long\def\epsfaux#1#2:#3\\{\ifx#1\epsfpercent
   \def\testit{#2}\ifx\testit\epsfbblit
      \epsfgrab #3 . . . \\%
      \epsffileokfalse
      \global\epsfbbfoundtrue
   \fi\else\ifx#1\par\else\epsffileokfalse\fi\fi}%
%
%
\def\epsfgrab #1 #2 #3 #4 #5\\{%
   \global\def\epsfllx{#1}\ifx\epsfllx\empty
      \epsfgrab #2 #3 #4 #5 .\\\else
   \global\def\epsflly{#2}%
   \global\def\epsfurx{#3}\global\def\epsfury{#4}\fi}%
%
%
\def\epsfsize#1#2{\epsfxsize}
%
%
\let\epsffile=\epsfbox

\begin{figure}
\epsfxsize=10cm
\hspace{2.5cm}\epsffile{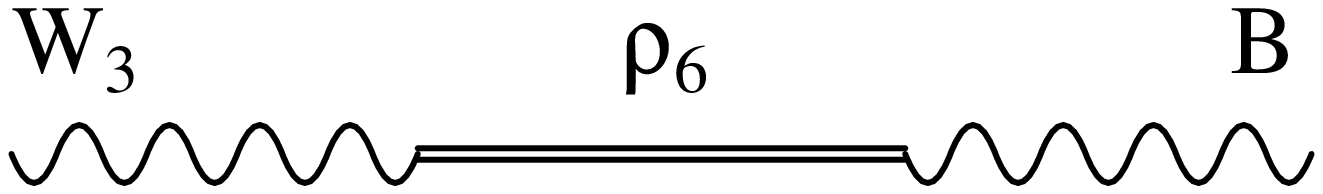}

\epsfxsize=10cm
\hspace{2.5cm}\epsffile{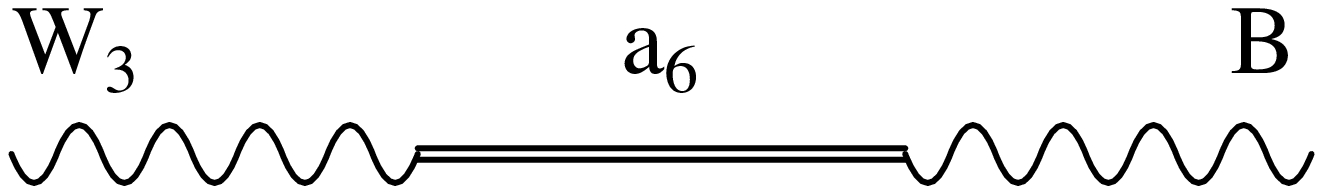}

\vspace{0.7cm}
\begin{center}
{\bf Fig.1}
\end{center}
\label{fig1}
\end{figure}
\medskip

\begin{figure}[htb]
\epsfxsize=10cm
\hspace{2.5cm}\epsffile{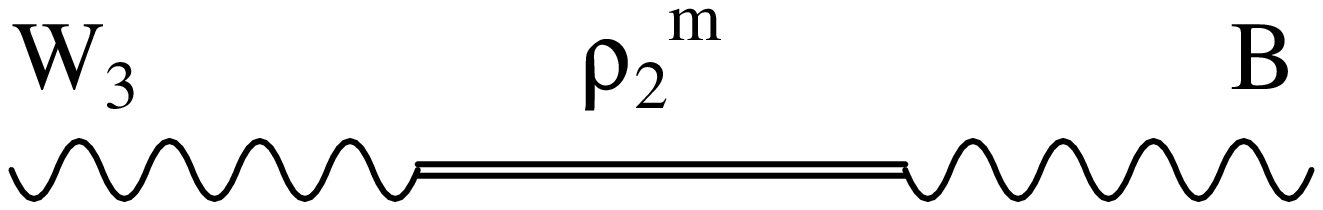}

\epsfxsize=10cm
\hspace{2.5cm}\epsffile{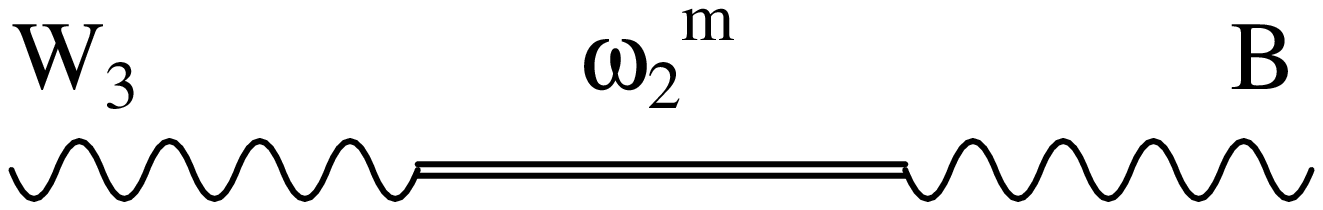}

\epsfxsize=10cm
\hspace{2.5cm}\epsffile{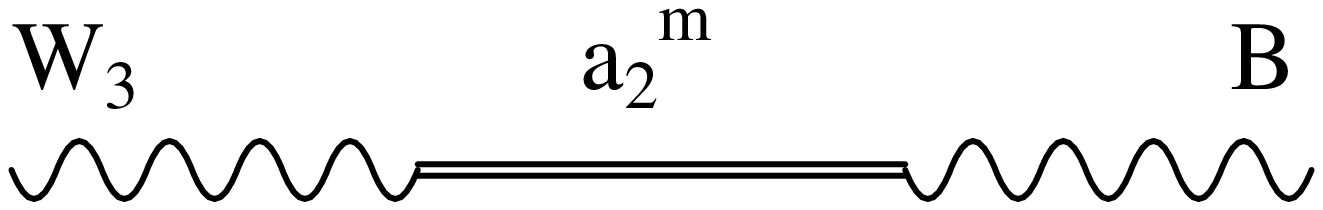}

\epsfxsize=10cm
\hspace{2.5cm}\epsffile{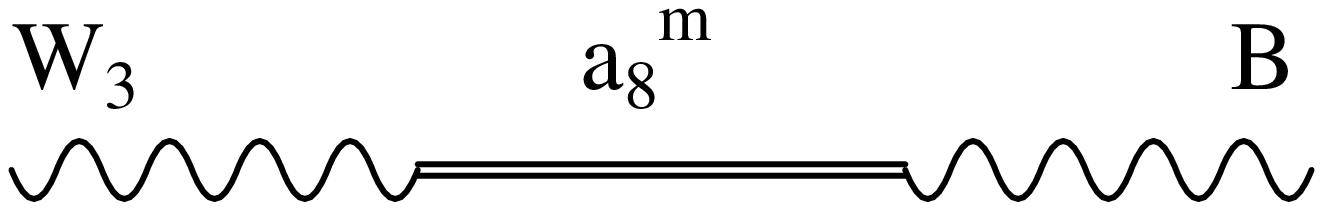}

\vspace{0.7cm}
\begin{center}
{\bf Fig.2}
\end{center}
\label{fig2}
\end{figure}

\begin{figure}
\epsfxsize=10cm
\hspace{2.5cm}\epsffile{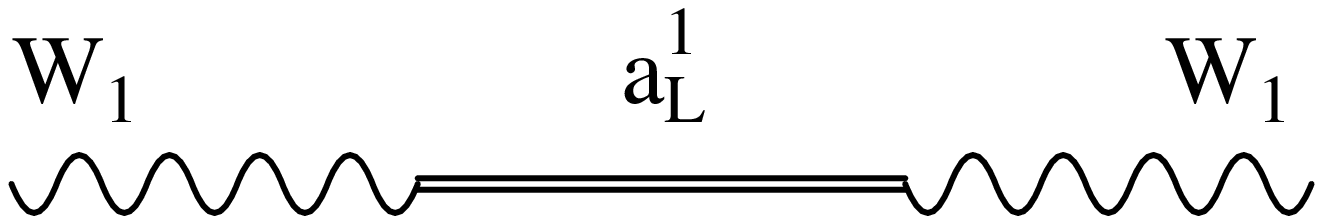}

\vspace{0.7cm}
\begin{center}
{\bf Fig.3}
\end{center}
\label{fig1}
\end{figure}
\medskip

\newpage
\begin{figure}
\setlength{\unitlength}{0.240900pt}
\begin{picture}(1500,900)(0,0)
\tenrm
\thicklines \path(220,113)(240,113)
\thicklines \path(1436,113)(1416,113)
\put(198,113){\makebox(0,0)[r]{0}}
\thicklines \path(220,266)(240,266)
\thicklines \path(1436,266)(1416,266)
\put(198,266){\makebox(0,0)[r]{1}}
\thicklines \path(220,419)(240,419)
\thicklines \path(1436,419)(1416,419)
\put(198,419){\makebox(0,0)[r]{2}}
\thicklines \path(220,571)(240,571)
\thicklines \path(1436,571)(1416,571)
\put(198,571){\makebox(0,0)[r]{3}}
\thicklines \path(220,724)(240,724)
\thicklines \path(1436,724)(1416,724)
\put(198,724){\makebox(0,0)[r]{4}}
\thicklines \path(220,877)(240,877)
\thicklines \path(1436,877)(1416,877)
\put(198,877){\makebox(0,0)[r]{5}}
\thicklines \path(220,113)(220,133)
\thicklines \path(220,877)(220,857)
\put(220,68){\makebox(0,0){0}}
\thicklines \path(423,113)(423,133)
\thicklines \path(423,877)(423,857)
\put(423,68){\makebox(0,0){1}}
\thicklines \path(625,113)(625,133)
\thicklines \path(625,877)(625,857)
\put(625,68){\makebox(0,0){2}}
\thicklines \path(828,113)(828,133)
\thicklines \path(828,877)(828,857)
\put(828,68){\makebox(0,0){3}}
\thicklines \path(1031,113)(1031,133)
\thicklines \path(1031,877)(1031,857)
\put(1031,68){\makebox(0,0){4}}
\thicklines \path(1233,113)(1233,133)
\thicklines \path(1233,877)(1233,857)
\put(1233,68){\makebox(0,0){5}}
\thicklines \path(1436,113)(1436,133)
\thicklines \path(1436,877)(1436,857)
\put(1436,68){\makebox(0,0){6}}
\thicklines \path(220,113)(1436,113)(1436,877)(220,877)(220,113)
\put(45,495){\makebox(0,0)[l]{\shortstack{$G_6/G_{\omega2}$}}}
\put(828,23){\makebox(0,0){$G_6/G_2$}}
\put(828,648){\makebox(0,0)[l]{\large{$S < 0$}}}
\put(828,266){\makebox(0,0)[l]{\large{$S > 0$}}}
\thinlines \path(283,877)(294,771)(306,686)(318,623)(331,575)
(343,537)(355,508)(367,484)(380,464)(392,448)(404,435)(417,423)
(429,414)(441,406)(453,400)(466,395)(478,390)(490,387)(503,384)
(515,382)(527,380)(539,379)(552,378)(564,378)(576,378)(588,378)
(601,379)(613,379)(625,380)(638,382)(650,383)(662,385)(674,387)
(687,388)(699,391)(711,393)(724,395)(736,398)(748,400)(760,403)
(773,405)(785,408)(797,411)(810,414)(822,417)(834,420)(846,423)
(859,427)(871,430)(883,433)(896,436)
\thinlines \path(896,436)(908,440)(920,443)(932,447)(945,450)
(957,454)(969,457)(982,461)(994,465)(1006,468)(1018,472)
(1031,476)(1043,480)(1055,483)(1068,487)(1080,491)(1092,495)
(1104,499)(1117,503)(1129,507)(1141,511)(1153,515)(1166,519)
(1178,523)(1190,527)(1203,531)(1215,535)(1227,539)(1239,543)
(1252,547)(1264,551)(1276,555)(1289,559)(1301,563)(1313,568)
(1325,572)(1338,576)(1350,580)(1362,584)(1375,588)(1387,593)
(1399,597)(1411,601)(1424,605)(1436,610)
\end{picture}

\begin{center}
{\bf Fig.4}
\label{Fig3}
\end{center}
\end{figure}


\begin{thebibliography}{99}
\bibitem{pes}
   B. Holdom and J. Terning, {\it Phys. Lett.} {\bf B247}
                                           (1990) 88 ;\\
   M. Golden and L. Randall, {\it Nucl. Phys.} {\bf B361}
                                           (1991) 3 ;\\
   M. E. Peskin and T. Takeuchi, {\it Phys. Rev. Lett.} {\bf
                                   65} (1990) 964 ;
                                {\it Phys. Rev.} {\bf D46} (1992) 381.
 \bibitem{sexp}
   R. Tanaka, Proceedings of XXVI International conference
on High Energy Physics, (1992) 681, Dallas, Texas, Edited by
J. R. Sanford.
 \bibitem{apel}
   T. Appelquist and J. Terning, {\it Phys. Lett.} {\bf B315} (1993) 139.
 \bibitem{wein}
   S. Weinberg, {\it Phys. Rev.} {\bf 166}, (1968) 1568.
 \bibitem{ban}
   M. Bando, T. Kugo, and K. Yamawaki,{\it Phys. Rep.}{\bf
164}
                          (1988) 217 ; and references therein.
  \bibitem{eck} G. Ecker, A. Pich, and E. de Rafael,
{\it Nucl. Phys.} {\bf B321} (1989) 311,
  J. F. Donoghue, and B. R. Holstein,
{\it Phys. Rev.} {\bf D40} (1989) 2378.   
  \bibitem{ono}
   T.Onogi and Y.Yasui, {\it Mod. Phys. Lett.}{\bf A7}(1992)
                                               785 ; \\ 
   C. T. Hill, D. C. Kennedy, T. Onogi and H. L. Yu, 
                                 {\it Phys. Rev.}{\bf D47} 
                                 (1993) 2940.
\end{thebibliography}
\end{document}